\documentclass[12pt]{article}
\usepackage{wider}
\begin{document}

\newcommand{\sqvb}{\ensuremath{ \langle \!\langle 0 |} }
\newcommand{\sqvk}{\ensuremath{ | 0 \rangle \!\rangle } }
\newcommand{\sqvn}{\ensuremath{ \langle \! \langle 0 |  0 \rangle \! \rangle} }
\newcommand{\wh}{\ensuremath{\widehat}}
\newcommand{\be}{\begin{equation}}
\newcommand{\ee}{\end{equation}}
\newcommand{\bea}{\begin{eqnarray}}
\newcommand{\eea}{\end{eqnarray}}
\newcommand{\ra}{\ensuremath{\rangle}}
\newcommand{\la}{\ensuremath{\langle}}
\newcommand{\rra}{\ensuremath{ \rangle \! \rangle }}
\newcommand{\lla}{\ensuremath{ \langle \! \langle }}
\newcommand{\str}{\rule[-.125cm]{0cm}{.5cm}}
\newcommand{\pr}{\ensuremath{^\prime}}
\newcommand{\ppr}{\ensuremath{^{\prime \prime}}}
\newcommand{\da}{\ensuremath{^\dag}}
\newcommand{\as}{^\ast}
\newcommand{\eps}{\ensuremath{\epsilon}}
\newcommand{\ve}{\ensuremath{\vec}}
\newcommand{\ka}{\kappa}
\newcommand{\non}{\ensuremath{\nonumber}}
\newcommand{\lf}{\ensuremath{\left}}
\newcommand{\rt}{\ensuremath{\right}}
\newcommand{\al}{\ensuremath{\alpha}}
\newcommand{\dfn}{\ensuremath{\equiv}}
\newcommand{\ga}{\ensuremath{\gamma}}
\newcommand{\ti}{\ensuremath{\tilde}}
\newcommand{\wti}{\ensuremath{\widetilde}}
\newcommand{\hs}{\ensuremath{\hspace*{.5cm}}}
\newcommand{\bet}{\ensuremath{\beta}}
\newcommand{\om}{\ensuremath{\omega}}

\newcommand{\cO}{\ensuremath{{\cal O}}}
\newcommand{\cS}{\ensuremath{{\cal S}}}
\newcommand{\cF}{\ensuremath{{\cal F}}}
\newcommand{\cX}{\ensuremath{{\cal X}}}
\newcommand{\cZ}{\ensuremath{{\cal Z}}}
\newcommand{\cG}{\ensuremath{{\cal G}}}
\newcommand{\cR}{\ensuremath{{\cal R}}}
\newcommand{\cV}{\ensuremath{{\cal V}}}
\newcommand{\cC}{\ensuremath{{\cal C}}}
\newcommand{\cP}{\ensuremath{{\cal P}}}
\newcommand{\pup}{\ensuremath{^{(p)}}}
\newcommand{\prpr}{\ensuremath{\prime \prime }}

\newcommand{\xxx}[1]{}
\newcommand{\zzz}{}

%citesupernumberparens.tex: Parentheses around superscript added by Mark A. Rubin 7/11/02
%rubin@ll.mit.edu
%
% version = 1.02 of citesupernumber.sty 1998 May 3
% origin 1989 January 20
% originally named citsupernumber.tex, but citesupernumber.sty works
%
% call in LaTeX2e:
% \usepackage{citesupernumber}
%
% This file makes citations be superscripted.  They come out as numbers.
% For example, by putting the following code in your LaTeX file,
%
%     text text text\cite{Darwin1859,Clark2001} text ...
%     \input citesupernumber
%     moretext moretext moretext\cite{Darwin1859,Clark2001} moretext ...
%
% you will get:
% 
%     text text text[1,2] text ...
%
%                              1,2
%     moretext moretext moretext   moretext ...

% NOTE:  By making this a package, the above switch does not work!!
% But then, would one really want to switch like that?

%
% This code was originally made by 
% Michael DeCorte // (315)265-2439 // P.O. Box 652, Potsdam, NY 13676
% Internet: mrd@sun.soe.clarkson.edu  // Bitnet:   mrd@clutx.bitnet        
% 
% It was modified by Tom Schneider toms@ncifcrf.gov

% DeCorte's original code was:
% \def\@cite#1#2{$^{#1\if@tempswa , #2\fi}$}
% I have added the \scriptsize so that the numbers are appropriately smaller
% and the mbox to make scriptsize work under the mathmode that is
% used to make the superscripting.
%

\makeatletter
\def\@cite#1#2{$^{\mbox{\scriptsize (#1\if@tempswa , #2\fi)}}$}
\makeatother

% ---------------------------------------------------
% If the following line:
% \def\cite#1{$^{#1}$}
% is also used, the output would be:
%                               Darwin1859,Clark2001
%     moretext moretext moretext                    moretext ...
%
% ---------------------------------------------------
% The next improvement of this would be to make it more like a
% switch, so one would say \citesupernumber or \citesuperkey
% or \citebracket or \citeparenthesis to get different forms.
% Better yet would be \citationform{place}{kind}{edge}
% where
%       place is 'inline' or 'superscript'
%       kind is 'citationkey' or 'number'
%       edge is 'bracket' or 'parenthesis' or 'none'
% Do you know how to make this happen?
% 
% While we are into wishing, it would be useful for some journals
% to have the following citation forms built, not from the key,
% but from the year and author list:
% One author:   (Smith, 1982)
% Two authors:  (Smith and Jones 1892)
% More than two authors:  (Smith {\em et al}, 1982)
% with ';' as separators between the citations.
% (For use in Journal of Molecular Biology.)

\title{\bf 
Spatial Degrees of Freedom in Everett Quantum Mechanics\thanks{
This work was sponsored by the Air Force under Air Force Contract 
FA8721-05-C-0002.  Opinions, interpretations, conclusions, and 
recommendations are those of the author and are not necessarily endorsed 
by the U.S. Government.
}
}
\author{
Mark A. Rubin\\
\mbox{}\\   %REMOVE FOR DOUBLE-SPACE  
Lincoln Laboratory\\ 
Massachusetts Institute of Technology\\  
244 Wood Street\\                         
Lexington, Massachusetts 02420-9185\\      
rubin@LL.mit.edu\\ 
\\
}
\date{\mbox{}}
\maketitle

\begin{abstract}

Stapp claims that, when spatial degrees of freedom are taken into
account, Everett quantum mechanics is ambiguous due to a ``core basis problem.''
To examine an aspect of this claim I generalize the ideal measurement model to include translational
degrees of freedom for both the measured system and the measuring apparatus. Analysis of this generalized model using the Everett interpretation in the Heisenberg picture shows that it makes unambiguous predictions for the possible results of measurements and their respective probabilities.  The presence of translational degrees of freedom for the measuring apparatus affects the probabilities of measurement
outcomes in the same way that a mixed state for the measured system would. Examination of a  measurement scenario involving several observers illustrates  the consistency of the model  with perceived spatial localization of the measuring apparatus.

\mbox{}

\noindent Key words: Everett interpretation, quantum mechanics, basis problem, Heisenberg picture

\end{abstract}

\section{Introduction} \label{sec_intro}

\subsection{Aim of the Present Paper}

The idea that the Everett interpretation$\mbox{}^{\mbox{\scriptsize(1-4)}}$\/   
\zzz 
of quantum mechanics contains an intrinsic
ambiguity regarding the description of the measurement process
has been present in the literature for decades.$\mbox{}^{\mbox{\scriptsize(5-9)}}$\/ 
\zzz 
This notion, variously  referred to as the ``basis ambiguity,'' the ``preferred basis problem,''
or simply  the ``basis problem,'' is generally presented in the context of models of
measuring devices and measured systems with discrete degrees of freedom.  In the  context of
such models  I have recently given an explicit demonstration,\cite{Rubin04}
following the approach  of DeWitt,\cite{DeWitt72,DeWitt98}  that there is no basis problem; i.e., that the Everett-interpretation description of measurement 
is in fact unambiguous.  

All real-world physical systems contain, however, not
just discrete degrees of freedom, but also  continuous spatial 
(translational and rotational) degrees of freedom.  In this paper
I extend  the analysis of the purely-discrete case\cite{Rubin04} 
with this fact in mind. Specifically, I present a model of
measurement situations in which both the measurement apparatus
and the observed system possess translational degrees of freedom,
and show that, despite this extension, the Everett description of 
measurement remains unambiguous. As in the purely-discrete case, the model
is  one of {\em isolated}\/ measurements; i.e., no role is played by
decoherence (see Ref. 13 \zzz and references therein).

\subsection{Stapp's Core Basis Problem}

Stapp\cite{Stapp02} claims that, in systems with spatial degrees of freedom such as those considered in the present paper, the basis problem may be present in aggravated form.  Quantum theory must satisfy the following two requirements:

\begin{description} 
\item[Outcome determination.] The formalism must unambiguously determine which specific outcomes observers will perceive to occur in a measurement situation, as well as the respective associated probabilities. The outcomes must form a discrete set, since ``the normal rules for extracting well defined probabilities from a 
quantum state require the specification, or singling out, of a 
{\it discrete set} (i.e., a denumerable set) of orthogonal subspaces, one 
for each of a set of alternative possible experientially distinguishable 
observations.''\cite{Stapp02}
\item[Localization.] The predictions of the formalism must be consistent with observers'
perception that  
measuring devices have well-defined spatial locations.
\end{description}

Stapp argues that  Everett quantum mechanics fails to satisfy
these requirements, and terms this failure the ``core basis problem:''

\begin{quote}

\small

The essential point is that if the universe has been evolving since the 
big bang in accordance with the Schr\"{o}dinger equation, 
then it must by now 
be an amorphous structure in which every device is a smeared-out cloud 
of a continuum of different possibilities.\ldots 
Due to the uncertainty principle, each particle 
would have had a tendency to spread out. Thus various particles with various 
momenta would have been able to combine and condense in myriads of ways into 
bound structures, including measuring devices, whose centers, orientations, 
and fine details would necessarily be smeared out over continua of 
possibilities.\ldots
But how can a particular discrete set of orthogonal 
subspaces be picked out from an amorphous continuum by the action of 
the Schr\"{o}dinger equation alone?\ldots
[The problem is that of] specifying
a discrete basis for the probability computations associated with our
apparently discrete (Yes or No) experiences, when the Schr\"{o}dinger equation 
generates continuous evolution of an ever-amorphous state.\cite{Stapp02}

\end{quote}

Stapp also argues that decoherence, which is sometimes invoked
to allow Everett quantum mechanics to satisfy the outcome-determination
requirement, can only do so at the expense of producing a theory
violating the localization requirement.
Since, as mentioned above, the model presented here does not make use of
decoherence,  I will not address this specific issue.  

Rather, I will show directly, for the case of a simple model of measurements with translational
degrees of freedom but without decoherence, that
Everett quantum mechanics does in fact satisfy both the outcome-determination
and localization requirements.

\subsection{Everett Quantum Mechanics in the Heisenberg Picture}

It proves convenient to perform the analysis working in the Heisenberg picture.
In this picture, the properties of physical objects at  time $t$\/
are determined jointly by time-dependent operators and a time-independent state
vector which encodes initial-condition information specified at an initial time
$t_{in} < t$\/.\footnote{The specific choice for $t_{in}$\/ is arbitrary, although in
any given situation some choices for $t_{in}$\/ may be more convenient than 
others.} So, interpretation of the formalism cannot be based, as it is in the Schr\"{o}dinger
picture, upon ``branching'' of the state vector during the course of the measurement
interaction---the Heisenberg-picture state vector does not ``branch'' or in any other way change its form.   
Rather, the operator which represents the degree of freedom of the measuring device 
relevant for its role as a measuring device 
evolves from a form which represents 
a ``ready state'' or ``state of ignorance'' to a  characteristic form which 
indicates that the degree of freedom in question has split into Everett copies, each of which 
possesses a definite value. (Measurement in Heisenberg-picture Everett quantum mechanics
has been described in Refs. 15, 16, and 10, and %\cite{Rubin01,Rubin03,Rubin04} This process 
is reviewed in Sec. \ref{Review} below.)

The possible outcomes for a measurement depend on the spectrum of the operator representing that particular degree of freedom.  If the spectrum of this operator is discrete, the possible
outcomes will be distinct. So, only operators with discrete spectra can correspond
to measurement results/states of awareness.  Such operators will be present
in systems of finite size; % and bounded energy; 
e.g., the operators corresponding
to  the energy associated with  relative motion of bound constituents, and functions of these
operators.

Probabilities can be assigned to these distinct outcomes using the same
rules which  would apply were the discrete degrees of freedom the only ones
present, since these rules make reference only to the operator representing the outcome,
not to any other operators pertaining to the measuring apparatus (such as its position in space).   
As far as measurement outcomes
recorded/perceived by a measuring device/observer are concerned, the spatial degrees of freedom are simply
``along for the ride.''

One may object that translational degrees of freedom are measured all the time,
and that it must be possible to assign them outcomes and probabilities in a consistent manner.
However, any measurement of such a continuous variable is performed by some 
finite-size 
 discrete-spectrum measuring device, and ultimately perceived by a finite-size 
discrete-spectrum brain. So, even when measuring a continuous degree of freedom, the measuring/observing system splits into Everett copies corresponding to discrete outcomes.

What about spatial localization of  macroscopic objects such as 
measuring devices and brains? Does such an object indeed  become, in the Everett interpretation,
a ``smeared-out cloud 
of a continuum of different possibilities?''  

In the first-quantized Heisenberg picture which we are employing here, the operator representing
a discrete ``internal'' degree of freedom, corresponding to the value obtained
by a measuring device, does not {\em have} a spatial location.  Position, in the first-quantized formalism, is an operator, and so cannot parameterize another operator.\footnote{In  Ref. 17 \zzz  
I employ the  field-theoretic Heisenberg-picture formalism, in which spatial position is a
c-number and  parameterizes operator-valued fields,  to show  the locality of information flow in entangled second-quantized systems.  However, only the measured systems are explicitly represented by field operators there, not the measuring instruments.} Therefore, localization must be demonstrated operationally; that is, by computing the results of measurements performed by several measuring devices, and showing that the results are consistent with the perception of definite positions for the measuring devices.  An example of this procedure will be presented in Sec. \ref{ConsistencyMultipleSpatial} below.  The upshot is that, provided the measurement {\em interactions}\/ are localized in space, the results of measurements are consistent with perceived spatial localization of the devices which perform the measurements.

\subsection{Organization of the Paper}

Sec. \ref{Review} reviews the formalism of ideal measurements in Heisenberg-picture
Everett quantum mechanics. 
Sec. \ref{Translation} 
introduces a model for finite-range interactions of measured systems and
measuring devices with translational degrees of freedom, analyzes this
model in the context of the Heisenberg-picture Everett interpretation, and shows that it makes
unambiguous predictions for the possible outcomes of measurements and their respective
probabilities.  The translational degrees of freedom of the measuring device, usually ignored,
are shown to have  the same effect on probabilities of measurement outcomes as would
an initial mixed state for the measured system. Sec. \ref{ConsistencyMultipleSpatial}
analyzes a measurement situation involving more than one observer, showing that
the formalism is consistent with the notion  
of measuring devices having well-defined
spatial locations. These results are summarized and discussed in  
Sec. \ref{Discussion}. The Appendix extends the operator expansion uniqueness theorem 
of Ref. 10
\zzz 
 to systems with continuous degrees of freedom.

\section{Heisenberg-Picture Everett Quantum Theory with Discrete Degrees of Freedom}
\label{Review}

The basic model for a measurement situation in quantum mechanics, so-called ``ideal measurement''  (see, e.g., Ch. 14 of Ref. 18\zzz), involves a physical system with a degree of freedom
$\cS$\/ the properties of which are measured by an observer with a degree of freedom $\cO$\/.  
These will generally be referred to simply as ``the system'' and ``the observer'' respectively.
(The use of the term ``observer'' is not meant to imply that $\cO$\/  is necessarily a degree
of freedom of a sentient system; e.g.,  $\cO\/$ might be a degree of freedom of a piece of laboratory apparatus.)
The respective state spaces of $\cS$\/ and $\cO$\/ are spanned by the vectors
\be 
|\cS;\al_i\ra, \hspace{.5cm}  i=1,\ldots, M, \label{basisS}
\ee
\be
|\cO;\bet_i\ra, \hspace{.5cm}  i=0,\ldots, M, \label{basisO}
\ee
where the labels $\al_i$\/ and $\bet_i$\/ are, respectively,  the nondegenerate eigenvalues of
Hermitian time-independent (i.e., Schr\"{o}dinger-picture) operators $\wh{a}$\/ and $\wh{b}$\/:
\be
\wh{a}|\cS;\al_i\ra=\al_i|\cS:\al_i\ra, \hspace*{.5cm}i=1,\ldots,M, \label{HPbasisS}
\ee
\be
\al_i = \al_j \Rightarrow i=j, \hspace*{.5cm}i,j=1,\ldots,M,\label{anondegen}
\ee
\be
\wh{b}|\cO;\bet_i\ra=\bet_i|\cO;\bet_i\ra, \hspace*{.5cm}i=0,\ldots,M,\label{HPbasisO}
\ee
\be
\bet_i = \bet_j \Rightarrow i=j, \hspace*{.5cm}i,j=0,\ldots,M.\label{bnondegen}
\ee
The vectors (\ref{basisS}), (\ref{basisO}) are taken to be normalized and so,
from (\ref{HPbasisS})-(\ref{bnondegen}), are orthonormal:
\be
\la \cS;\al_i|\cS;\al_j \ra = \delta_{ij},\hspace*{.5cm} i,j=1,\ldots,M, \label{Sorthon}
\ee
\xxx{Sorthon}
\be
\la \cO;\bet_i|\cO;\bet_j \ra = \delta_{ij},\hspace*{.5cm} i,j=0,\ldots,M. \label{Oorthon}
\ee
\xxx{Oorthon}

The measurement is a physical interaction between $\cS$\/ and $\cO$\/.  Prior to
the measurement the combined state of $\cS$\/ and $\cO$\/ is a product state,
implying that the respective observables of $\cS$\/ and $\cO$\/  
are uncorrelated.  In addition, $\cO$\/ is taken
to be in the ``ignorant'' or ``ready''  state $|\cO;\bet_0\ra$\/ while the state
of $\cS$\/ is arbitrary.  It is convenient
to take the time immediately prior to the measurement to be the time $t_{in}$\/
at which the time-independent Heisenberg-picture state vector $|\psi(t_{in})\ra$\/
is defined. That is, $t_{in}$\/ is the time at which the Heisenberg-picture state vector
is equal to  the  Schr\"{o}dinger-picture state vector.
Then
\be
|\psi(t_{in})\ra=|\cO;\bet_0\ra |\cS;\psi\ra,   \label{initialHPstate}
\ee
\xxx{initialHPstate}
where
\be
|\cS;\psi\ra = \sum_{i=1}^M \psi_i |\cS;\al_i\ra , \label{Sexpansion}
\ee
\be
\sum_{i=1}^M |\psi_i|^2=1, \hspace{.5cm}  {\hbox{\rm $\{\psi_i\}$\/ otherwise arbitrary.}} \label{psiarb}
\ee  

Between time $t_{in}$\/ and time $t$\/, any operator  $\wh{o}$\/ evolves according
to the Heisenberg-picture dynamical rule
\be
\wh{o}(t)=\wh{U}_I\da \wh{o}(t_{in}) \wh{U}_I. \label{UactionHP}
\ee
where $\wh{U}_I$\/ is the unitary time-evolution operator,
\be
\wh{U}_I=\exp(-i\wh{H}_I(t-t_{in})).\label{UI}
\ee
We will generally drop the time argument in Heisenberg-picture operators
evaluated at the initial time $t_{in}$\/, since they are then just equal
to their Schr\"{o}dinger-picture counterparts:
\be
\wh{o}(t_{in})=\wh{o}.
\ee
For an ideal measurement, the Hamiltonian $\wh{H}_I$\/  has  the 
form 
\be
\wh{H}_I=\sum_{i=1}^M \wh{h}_i^\cO \otimes \wh{P}_i^\cS.\label{HI}
\ee
Here  $\wh{h}_i^\cO$\/ acts only in the space of $\cO$\/,  
and
$\wh{P}_i^\cS$\/ is the projection operator into the $i^{\rm th}$\/ to-be-measured state of $\cS$\/:
\be
\wh{P}_i^\cS=|\cS;\al_i \ra \la \cS;\al_i |, \hspace{5mm} i=1,\ldots,M, \label{Sproj}
\ee
\xxx{Sproj}
satisfying
\be
\wh{P}_i^{\cS}\wh{P}_j^{\cS}=\delta_{ij}\wh{P}_j^{\cS},\label{PcSortho}
\ee
\be
\sum_{i=1}^M\wh{P}_i^\cS=1.\label{PcScomplete}
\ee
\xxx{PcScomplete}
Using (\ref{Sorthon}), (\ref{HI}) and (\ref{Sproj}) in (\ref{UI}),
\be
\wh{U}_I=\sum_{i=1}^M \wh{u}_i^\cO \otimes \wh{P}_i^\cS,\label{UIform}
\ee
where
\be
\wh{u}_i^\cO=\exp(-i\wh{h}_i^\cO (t-t_{in})), \hspace*{5mm} i=1,\ldots,M.\label{uiform}
\ee
For an ideal measurement,
the
$\wh{u}_i^\cO$\/ must have the property that
\be
\wh{u}_i^\cO |\cO;\bet_0\ra=|\cO;\bet_i\ra, \hspace{5mm}i=1,\ldots,M.  \label{uiO}
\ee
E.g., if the $\wh{h}_i^\cO$\/'s are of the form
\be
\wh{h}_i^{\cO}=i\kappa \left(\;|\cO;\bet_i\ra\la \cO;\bet_0|-|\cO;\bet_0\ra\la \cO;\bet_i|\; \right),\hspace*{5mm}i=1,\ldots,M,
\label{particular_h}
\ee 
then (\ref{uiO}) will be satisfied provided
\be
\kappa=\frac{\pi}{2(t-t_{in})}.
\ee
From (\ref{Sorthon}), (\ref{UactionHP}),  
(\ref{Sproj})  and   
(\ref{UIform}) it follows that 
\be
\wh{b}(t)=\sum_{i=1}^M \wh{b}_i \otimes \wh{P}_i^\cS, \label{b_t_expansion}
\ee
where
\be
\wh{b}_i=\wh{u}_i^{\cO\dagger} \wh{b} \wh{u}_i^\cO, \hspace*{5mm}i=1,\ldots,M. \label{bidef}
\ee
\xxx{bidef}
From (\ref{HPbasisO}), (\ref{uiO}) and (\ref{bidef}),
\be
\wh{b}_i|\cO;\bet_0\ra=\bet_i|\cO;\bet_0\ra \hspace*{5mm} i=1,\ldots,M. \label{b_i_eigenvalue}
\ee
\xxx{b-i-eigenvalue}

The fact that $\wh{b}(t)$\/ has the form (\ref{b_t_expansion}) and satisfies (\ref{b_i_eigenvalue}) with the initial state of the form (\ref{initialHPstate}) is 
taken to indicate that  the physical degree of freedom $\cO$\/ described by $\wh{b}(t)$\/ has
been split into $M$\/ Everett copies with respective values $\beta_1,\ldots,\beta_M$\/. 
 This is termed
``interpretive rule 1.'' \cite{Rubin03}

As has been shown in Sec. 4.2.1 of Ref. 10, \zzz 
the decomposition (\ref{b_t_expansion}) satisfying (\ref{b_i_eigenvalue}) is unique. 
I.e.,   
there is no inequivalent decomposition of $\wh{b}(t)$\/ 
of the form (\ref{b_t_expansion}) satisfying (\ref{b_i_eigenvalue}) 
which would  
correspond to the observer having measured some different property of the system.
There is thus no ``basis ambiguity.''

The projectors
$\wh{P}^\cS_i$\/ which multiply the $\wh{b}_i$\/'s
serve as ``labels''\cite{Rubin01} attached to
the Everett copies, carrying information about past interactions  and playing an important in role in the Everett-interpretation resolution of
the EPR paradox.\cite{Tipler00,DeutschHayden00,Rubin01,Rubin02}

Taken by itself, interpretive rule 1 would have us conclude that all physically-possible outcomes to a measurement are realized every time that measurement is performed, even when the amplitudes $\psi_i$\/ for
one or more outcomes are equal to zero.  To enable the formalism to be applicable to situations in which not all possible outcomes are realized,   interpretive rule 1 must be supplemented\cite{Rubin03} by 
interpretive rule 2: Only those copies of $\cO$\/ exist at time $t$\/ which 
satisfy $W_i(t) \neq 0$\/, where the ``weight'' $W_i(t)$\/ associated with the $i^{\rm th}$\/ copy is the matrix element  of the label between the initial-state bra and ket:
\be
W_i(t)=\la \psi(t_{in})|\wh{P}^{\cS}_i|\psi(t_{in})\ra. 
\label{weight_formula}
\ee

By introducing an ensemble of systems identical to $\cS$\/ and accompanied
by their respective observers $\cO$\/, one can, using interpretive rules 1 and 2
and physically-based restrictions on the properties of measurement devices
(finite resolution and information-storage capacity), {\em derive}\/\cite{Rubin03}   both the existence of probability in quantum mechanics and the Born rule for the value of probability, 
where probability is in the sense of the frequency 
interpretation, i.e,  relative frequency in the limit of an infinitely-large ensemble(see, e.g., Refs. 21 and 22.\zzz)
The use of the Heisenberg picture avoids problems associated
with other approaches to deriving frequency-interpretation probability in the Everett
interpretation (see Ref. 16 \zzz 
for details and references).
The value of the probability associated with the 
$i^{\rm th}$\/ outcome
of the ideal measurement interaction  
is $W_i(t)$\/; using (\ref{PcScomplete}) in (\ref{weight_formula}), we see
that
\be
\sum_{i=1}^M W_i(t)=\la \psi(t_{in})|\sum_{i=1}^M\wh{P}^{\cS}_i|\psi(t_{in})\ra=1.
\label{unitarity}
\ee
\xxx{unitarity}

\section{Heisenberg-Picture Everett Quantum Theory with Translational Degrees of Freedom}
\label{Translation}
\xxx{Translation}

\subsection{A Localized Measurement Interaction Model}\label{LocalMeasurement}

\subsubsection{State Spaces}\label{Translationmodelstate}
\xxx{Translationmodelstate}

To investigate the role of spatial degrees of freedom, we must extend the  ideal-measure\-ment model.  We first introduce translational degrees of freedom $\cal{X}$\/ and $\cal{Z}$\/ for the system and the observer, respectively. The state space is spanned by tensor products of the discrete sets of basis vectors (\ref{basisS}), (\ref{basisO}) and the continuous sets of basis vectors
\be
|\cX_i;\xi_i\ra, \hspace*{5mm}-\infty < \xi_i < \infty, \hspace*{5mm}i=1,2,3, \label{cXivec}
\ee
\xxx{cXivec}
\be
|\cZ_i;\zeta_i\ra, \hspace*{5mm}-\infty < \zeta_i < \infty, \hspace*{5mm}i=1,2,3,\label{cZivec}
\ee
where the labels $\xi_i$\/, $\zeta_i$ are three-dimensional spatial coordinates, eigenvalues of the corresponding position operators:
\be
%\mbox{}^{\;\;}
\wh{\vec{x}}|\cX;\vec{\xi}\:\ra=\vec{\xi}|\cX;\vec{\xi}\:\ra,  \label{xevalevec}
\ee
\be
\wh{\vec{z}}|\cZ;\vec{\zeta}\:\ra=\vec{\zeta}|\cZ;\vec{\zeta}\:\ra,
\ee
where
\be
\wh{\vec{x}}=(\wh{x}_1,\wh{x}_2,\wh{x}_3),\label{xhatvec}
\ee
\be
\wh{\vec{z}}=(\wh{z}_1,\wh{z}_2,\wh{z}_3),
\ee
\be
\vec{\xi}=(\xi_1,\xi_2,\xi_3),\label{xivec}
\ee
\be
\vec{\zeta}=(\zeta_1,\zeta_2,\zeta_3),\label{zetavec}
\ee
\be
|\cX;\vec{\xi} \: \ra=\bigotimes_{i=1}^3 |\cX_i;\xi_i\ra, \label{cXvec}
\ee
\be
|\cZ;\vec{\zeta} \: \ra=\bigotimes_{i=1}^3 |\cZ_i;\zeta_i\ra.\label{cZvec}
\ee
In terms of the vectors (\ref{cXivec}), (\ref{cZivec})
(\ref{cXvec}) and (\ref{cZvec}), we can define the $\cX$\/ and  $\cZ$\/ projection operators 
\be
\wh{P}_{\vec{\xi}}^{\cX}=\bigotimes_{i=1}^3 \wh{P}_{\xi_i}^{\cX_i}=|\cX;\vec{\xi}\ra \la \cX; \vec{\xi}|,  \label{PxicX}
\ee
\be 
\wh{P}_{\vec{\zeta}}^{\cZ}=\bigotimes_{i=1}^3\wh{P}_{\zeta_i}^{\cZ_i}=|\cZ;\vec{\zeta}\ra \la \cZ_;\vec{\zeta}|, \label{PzetacZ}
\ee
where
\be 
\wh{P}_{\xi_i}^{\cX_i}=|\cX_i;\xi_i\ra \la \cX_i;\xi_i|,\hspace*{5mm}i=1,2,3,
\ee
\be 
\wh{P}_{\zeta_i}^{\cZ_i}=|\cZ_i;\zeta_i\ra \la \cZ_i;\zeta_i|,\hspace*{5mm}i=1,2,3.
\label{PzetaicZi}
\ee
\xxx{PzetaicZi}

In the remainder of this paper we will occasionally use the symbols
$\cS$ and $\cO$\/ to refer, not just to the  discrete degrees
of freedom, but to the measured system and observer as a whole, including
the  spatial degrees of freedom $\cX$\/ and $\cZ$\/.
Which meaning is used in any specific case should be clear from
the context.

\subsubsection{Measurement Interaction}\label{MeasIntsubsubsec}

We next introduce a spatially-localized measurement interaction.  Define the operator 
\be
\wh{P}_f^{\cZ \cX}=\int d^3\vec{\zeta}\:d^3\vec{\xi}\: f(\vec{\zeta},\vec{\xi})
	\wh{P}_{\vec{\zeta}}^{\cZ} \wh{P}_{\vec{\xi}}^{\cX}, \label{PfcZcX}
\ee
where $f(\vec{\zeta},\vec{\xi})$\/ is a c-number function, and   require that $\wh{P}_f^{\cZ \cX}$\/ be a projection operator,
\be
\left({\wh{P}_f^{\cZ \cX}}\right)^2=\wh{P}_f^{\cZ \cX}.\label{PcZcXprojcond}
\ee
Using (\ref{PxicX})-(\ref{PfcZcX}) and imposing the usual normalization conditions for $|\cX;\vec{\xi}\ra$\/, $|\cZ;\vec{\zeta}\ra$\/,
\be
\la\cX;\vec{\xi}|\cX;\vec{\xi}\pr\ra=\delta^3(\vec{\xi}-\vec{\xi}\pr),
\label{cXorthog}
\ee
\be
\la\cZ;\vec{\zeta}|\cZ;\vec{\zeta}\pr\ra=\delta^3(\vec{\zeta}-\vec{\zeta}\pr),
\label{cZorthog}
\ee
(\ref{PcZcXprojcond}) implies that 
\be
\left(f(\vec{\zeta},\vec{\xi})\right)^2=f(\vec{\zeta},\vec{\xi}) \hspace*{5mm} \forall \; \vec{\zeta},\vec{\xi},
\ee
i.e.,
\be
f(\vec{\zeta},\vec{\xi})= \mbox{\rm 0 or 1} \hspace*{5mm} \forall \; \vec{\zeta},\vec{\xi}.
\label{f0or1}
\ee

To model measurement interactions of finite range, we choose
\be
f(\vec{\zeta},\vec{\xi})=\theta(a-|\vec{\zeta}-\vec{\xi}|), \hspace*{5mm}a > 0,\label{f_form}
\ee
where $\theta$\/ is the Heaviside unit step function. With the choice  (\ref{f_form}) for $f$\/, $\wh{P}_f^{\cZ\cX}$\/ is the
projection operator into the subspace of states in $\cZ \otimes \cX$\/ which are within 
spatial distance $a$\/ of each other:
\be
\begin{array}{rll}
\wh{P}_f^{\cZ\cX}|\cZ;\vec{\zeta}\ra|\cX;\vec{\xi}\ra&=|\cZ;\vec{\zeta}\ra|\cX;\vec{\xi}\ra,
\hspace*{5mm}& |\vec{\zeta}-\vec{\xi}| < a,\\
\mbox{}&=0, \hspace*{2.3cm}& |\vec{\zeta}-\vec{\xi}| > a.
\end{array} \label{PcZcXproj}
\ee

The finite-range measurement Hamiltonian is taken to be
\be
\wh{H}_F=\wh{H}_I \otimes \wh{P}_f^{\cZ\cX}.\label{HF}
\ee
\xxx{HF}
where $\wh{H}_I$\/ is the ideal-measurement Hamiltonian (\ref{HI}). 
Taking into account (\ref{PcZcXproj}), the form (\ref{HF}) for $\wh{H}_F$\/ implies  that $\cS$\/ and $\cO$\/ only interact if they are within distance $a$\/ of each other.

Using (\ref{HI}), 
(\ref{PcSortho}), (\ref{uiform}), (\ref{PcZcXprojcond}), and (\ref{HF}), the time-evolution operator
\be
\wh{U}_F=\exp(-i\wh{H}_F(t-t_{in}))
\ee
has the form
\be
\wh{U}_F=\wh{P}_{\wti{f}}^{\cZ\cX}+\sum_{i=1}^M\wh{u}_i^{\cO}\otimes
\wh{P}_i^{\cS}\otimes\wh{P}_f^{\cZ\cX}.\label{UFform}
\ee
Here
\be
\wti{f}(\vec{\zeta},\vec{\xi})=1- f(\vec{\zeta},\vec{\xi}),\label{ftilde_form}
\ee
\be
\wh{P}_{\wti{f}}^{\cZ\cX}= \int d^3\vec{\zeta}d^3\vec{\xi}\; \wti{f}(\vec{\zeta},\vec{\xi})
	\;\wh{P}_{\vec{\zeta}}^{\cZ} \wh{P}_{\vec{\xi}}^{\cX}. \label{PftildecZcX}
\ee
$\wh{P}_{\wti{f}}^{\cZ\cX}$\/ is the projector into the subspace of states in $\cZ \otimes \cX$\/
which are farther apart than $a$\/:
\be
\begin{array}{rll}
\wh{P}_{\wti{f}}^{\cZ\cX}|\cZ;\vec{\zeta}\ra|\cX;\vec{\xi}\ra&=|\cZ;\vec{\zeta}\ra|\cX;\vec{\xi}\ra,
\hspace*{5mm}& |\vec{\zeta}-\vec{\xi}| > a,\\
\mbox{}&=0, \hspace*{2.3cm}& |\vec{\zeta}-\vec{\xi}| < a.
\end{array} \label{PcZcXtildeproj}
\ee
satisfying, from (\ref{PfcZcX}), (\ref{PcZcXprojcond}), (\ref{f0or1}), (\ref{ftilde_form}) and (\ref{PftildecZcX}),
\be
\left(\wh{P}_{\wti{f}}^{\cZ\cX}\right)^2=\wh{P}_{\wti{f}}^{\cZ\cX},\label{Pwtisq}
\ee
\be
\wh{P}_{\wti{f}}^{\cZ\cX}=1-\wh{P}_f^{\cZ\cX},\label{PZXfteq1mPZXf}
\ee
\xxx{PZXfteq1mPZXf}
\be
\wh{P}_{\wti{f}}^{\cZ\cX}\wh{P}_f^{\cZ\cX}=\wh{P}_f^{\cZ\cX}\wh{P}_{\wti{f}}^{\cZ\cX}=0.\label{PwtifPf}
\ee

\subsubsection{Time Evolution of Operators}

An operator which at time $t_{in}$\/ has the value $\wh{o}$\/ has, at the later time 
$t$\/, the value
\be
\wh{o}(t)=\wh{U}_F\da \;\wh{o} \; \wh{U}_F. \label{o_t}
\ee
If $\wh{d}=\wh{d}(t_{in})$\/ is an operator which at time $t_{in}$\/ acts nontrivially only in $\cO$\/-space, % i.e.
then, using (\ref{PcSortho}), (\ref{PcZcXprojcond}), (\ref{UFform}), (\ref{Pwtisq}), 
(\ref{PwtifPf}) and (\ref{o_t}),
\be
\wh{d}(t)=\wh{d}_0 \otimes \wh{P}_{\wti{f}}^{\cZ\cX}+
\sum_{i=1}^M \wh{d}_i\otimes \wh{P}_i^{\cS}\otimes \wh{P}_f^{\cZ\cX},
\ee
where
\be
\wh{d}_0=\wh{d},
\ee
\be
\wh{d}_i={\wh{u}_i^{\cO}}\mbox{}\da \: \wh{d} \: \wh{u}_i^{\cO}, \hspace*{5mm} i=1,\ldots,M.
\ee
In particular,
\be
\wh{b}(t)=\wh{b}_0 \otimes \wh{P}_{\wti{f}}^{\cZ\cX}+
\sum_{i=1}^M \wh{b}_i\otimes \wh{P}_i^{\cS}\otimes \wh{P}_f^{\cZ\cX},\label{b_t}
\ee
\xxx{b-t}
where
\be
\wh{b}_0=\wh{b}, \label{b_0}
\ee
and where the $\wh{b}_i$\/'s for $i=1,\ldots,M$\/ are as given in (\ref{bidef}).
The coefficients of the $\wh{b}_i$\/'s in (\ref{b_t}) are a complete
orthogonal set of projection operators:
\be
\left(\wh{P}_{\wti{f}}^{\cZ\cX}\right)^2=\wh{P}_{\wti{f}}^{\cZ\cX},%\label{Pwtisq}
\ee
\be
\wh{P}_{\wti{f}}^{\cZ\cX} \left( \wh{P}_i^{\cS} \otimes \wh{P}_f^{\cZ\cX} \right)
=\left( \wh{P}_i^{\cS} \otimes \wh{P}_f^{\cZ\cX} \right)\wh{P}_{\wti{f}}^{\cZ\cX} =0,
\hspace*{5mm}i=1,\ldots,M,
\ee
\be
\left( \wh{P}_i^{\cS} \otimes \wh{P}_f^{\cZ\cX} \right)
\left( \wh{P}_j^{\cS} \otimes \wh{P}_f^{\cZ\cX} \right)=\delta_{ij}
\left( \wh{P}_i^{\cS} \otimes \wh{P}_f^{\cZ\cX} \right),\hspace*{5mm}i,j=1,\ldots,M,
\ee
\be
\wh{P}_{\wti{f}}^{\cZ\cX}+\sum_{i=1}^M\left( \wh{P}_i^{\cS} \otimes \wh{P}_f^{\cZ\cX} \right)=1.
\ee
As in the completely-discrete case, $\wh{a}$\/ is unchanged by the measurement: 
\be
\wh{a}(t)=\wh{U}\da_F\;\wh{a}\;\wh{U}_F=\wh{a}.
\ee
The 
same is true for $\wh{\vec{x}}$\/ and $\wh{\vec{z}}$\/:
\be
\wh{\vec{x}}(t)=\wh{U}\da_F\;\wh{\vec{x}}\;\wh{U}_F=\wh{\vec{x}},
\ee
\be
\wh{\vec{z}}(t)=\wh{U}\da_F\;\wh{\vec{z}}\;\wh{U}_F=\wh{\vec{z}}.
\ee

\subsubsection{Initial State}

For a measurement situation, the initial state should be one in which the degrees of
freedom pertaining to the observer, $\cO$\/ and $\cZ$\/, are uncorrelated with those pertaining to the system to be measured, $\cS$\/ and $\cX$\/. In addition, a measurement of the $\cO$\/ degree of freedom should with certainty determine that the observer is in a state of ignorance.  We therefore take the initial state to be 
\be
|\psi_1(t_{in})\ra=|\cZ,\cO, t_{in}\ra |\cX,\cS, t_{in}\ra,\label{psi_t_in}
\ee
where
\be
|\cX,\cS, t_{in}\ra=\sum_{i=1}^M\int d^3\vec{\xi}\: \psi_i^{\cX\cS}(\vec{\xi})\:|\cX;\vec{\xi}\ra|\cS;\al_i\ra,\label{cXcStin}
\ee
\xxx{cXcStin}
\be
|\cZ,\cO, t_{in}\ra=\int d^3 \vec{\zeta} \: \psi^\cZ(\vec{\zeta})\: |\cZ;\vec{\zeta}\ra|\cO;\bet_0\ra,\label{cZcOtin}
\ee
\xxx{cZcOtin}
and where $\psi_i^{\cX\cS}(\vec{\xi})$\/ and $\psi^\cZ(\vec{\zeta})$\/ are
c-number functions. Normalization of $|\cX,\cS, t_{in}\ra$\/ and $|\cZ,\cO, t_{in}\ra$\/
imposes the constraints
\be
\sum_{i=1}^M \int d^3\vec{\xi}\:|\psi_i^{\cX\cS}(\vec{\xi})|^2=1,\label{normpsiXSi}
\ee
\xxx{normpsiXSi}
\be
\int d^3 \vec{\zeta} \: |\psi^\cZ(\vec{\zeta})|^2=1.\label{normpsiZ}
\ee
\xxx{normpsiZ}

Using 
(\ref{HPbasisO}), (\ref{b_i_eigenvalue})  
and (\ref{b_0}), 
we see that 
\be
\wh{b}_i|\cO;\beta_0\ra=\bet_i|\cO;\beta_0\ra,\hspace*{5mm}i=0,\ldots,M.\label{b_i_eigen2}
\ee
So, by virtue of interpretive rule 1, we conclude from (\ref{b_t}),
(\ref{psi_t_in}) and (\ref{b_i_eigen2}) that at time $t$\/ the observer, in a state of ignorance at time $t_{in}$\/,  
has split into $M+1$\/ Everett copies, one of which has remained in a state of ignorance, 
the others of which have respectively observed the $M$\/ possible outcomes of successful measurement of $\cS$\/.

Expression (\ref{b_t}) is the extension of the  expansion (\ref{b_t_expansion}) to a situation involving translational degrees of freedom. Is there any sort of basis ambiguity present here?
Could $\wh{b}(t)$\/ also be expanded in a manner different from and physically inequivalent
to (\ref{b_t})? The answer to these questions is ``no,'' 
as may be seen from the generalized version of the  operator
expansion uniqueness theorem of Sec. 4.2.1 of Ref. 10 \zzz 
that is derived in the Appendix of the present paper.

\subsubsection{Probability}\label{ProbabilitySingleObserver}

The probability associated with the Everett copy of $\cO$\/ which remains in a state
of ignorance is
\be
W_{b,0}=\la\psi_1(t_{in})|\wh{P}_{\wti{f}}^{\cZ\cX}|\psi_1(t_{in})\ra.
\ee

It should be  emphasized that the derivation of ``weight equals probability'' from 
relative frequency considerations has only to date been carried out  for
quantum systems with only discrete degrees of freedom.\cite{Rubin03} So,
for the present, its application here must  be considered to be  an independent assumption
(as is the usual practice in Everett and non-Everett quantum mechanics).

Using (\ref{PftildecZcX}) and (\ref{psi_t_in})-(\ref{cZcOtin})  
we find
\be
W_{b,0}=\sum_{i=1}^M\int d^3\vec{\zeta}\:d^3\vec{\xi}|\psi^\cZ(\vec{\zeta})|^2\:
\wti{f}(\vec{\zeta},\vec{\xi}) \: |\psi_i^{\cX\cS}(\vec{\xi})|^2. \label{Wb0}
\ee
The probabilities for the other Everett copies are
\be
W_{b,i}=\la \psi_1(t_{in})|\wh{P}_i^\cS \otimes \wh{P}_f^{\cZ\cX}|\psi_1(t_{in})\ra,
\hspace*{5mm}i=1,\ldots,M.
\ee
Using (\ref{Sorthon}), (\ref{Sproj}), (\ref{PfcZcX}) and (\ref{psi_t_in})-(\ref{cZcOtin})
\be 
W_{b,i}=\int d^3\vec{\zeta}\:d^3\vec{\xi}\:|\psi^\cZ(\vec{\zeta})|^2 
f(\vec{\zeta},\vec{\xi}) \: |\psi_i^{\cX\cS}(\vec{\xi})|^2,\hspace*{5mm}i=1,\ldots,M.
\label{Wbi}
\ee
\xxx{Wbi}
Using (\ref{PcScomplete}), (\ref{normpsiXSi}), (\ref{normpsiZ}), (\ref{Wb0}) and
(\ref{Wbi}), we verify that these probabilities properly sum to one:
\be
W_{b,0} + \sum_{i=1}^M W_{b,i}=\int d^3\vec{\zeta}\:d^3\vec{\xi}\:|\psi^\cZ(\vec{\zeta})|^2 
\left(f(\vec{\zeta},\vec{\xi})+ \wti{f}(\vec{\zeta},\vec{\xi})\right)\: 
\sum_{i=1}^M|\psi_i^{\cX\cS}(\vec{\xi})|^2=1.\label{sdofunitarity}
\ee
\xxx{sdofunitarity}

The Heisenberg-picture Everett formalism is thus seen to provide  unambiguous
outcome determination, even when the translational degrees of freedom
of the observer/measuring device are included in the description of the measurement situation.  I.e., the formalism specifies {\em what}\/ the outcomes to measurement are---including here the possibility that, due to excessive spatial separation between $\cS$\/ and $\cO$\/, no 
measurement is made---as well as the respective probabilities of the outcomes, (\ref{Wb0}) and (\ref{Wbi}).

\subsection{Observer Spatial Degrees of Freedom Mimic A Proper Mixture}

No restriction has been placed on the initial probability amplitude $\psi^\cZ(\vec{\zeta})$\/ for the translational degrees of freedom of the observer.  How it comes about that there is nevertheless no
 conflict with the observed spatial localization of macroscopic objects such 
as measuring devices will be examined in the following section.  Here we address a potential
objection based on the form of the results (\ref{Wb0}), (\ref{Wbi}).  Namely, these expressions for the probability of measurement outcomes explicitly involve the measuring-device-position amplitude $\psi^\cZ(\vec{\zeta})$\/.  Now, typically, when computing the respective probabilities of the possible  outcomes of measurement of a quantum system performed using a macroscopic measuring device, one  makes use of the  state vector of the measured system; but one does not have to take into account any sort of wavefunction for the measuring device itself. 

This apparent discrepancy is resolved by noting that the results (\ref{Wb0}), (\ref{Wbi}) of the Heisenberg-picture formalism  are equal to those obtained   
if one proceeds in the usual manner---i.e., ignoring spatial degrees of freedom of the measuring apparatus---and  computes the probabilities of measurement outcomes for a measured system initially in a {\em mixed state}\/.
Since, from (\ref{f_form}) and (\ref{ftilde_form}), 
$f(\vec{\zeta},\vec{\xi})$\/ and $\wti{f}(\vec{\zeta},\vec{\xi})$\/ are translation invariant,
\be
f(\vec{\zeta},\vec{\xi})=f(\vec{\zeta}-\vec{q},\vec{\xi}-\vec{q}) \hspace*{5mm}\forall \vec{q},
\label{f_invariant}
\ee
\be
\wti{f}(\vec{\zeta},\vec{\xi})=\wti{f}(\vec{\zeta}-\vec{q},\vec{\xi}-\vec{q}) \hspace*{5mm}\forall \vec{q},
\label{ftilde_invariant}
\ee
(\ref{Wb0}), (\ref{Wbi}) can be written as
\be
W_{b,0}=\sum_{i=1}^M\int d^3\vec{\zeta}\:d^3\vec{\xi}|\psi^\cZ(\vec{\zeta})|^2\:
\wti{f}(\vec{0},\vec{\xi}) \: |\psi_i^{\cX\cS}(\vec{\xi}+\vec{\zeta})|^2, \label{Wb0ti}
\ee
\be 
W_{b,i}=\int d^3\vec{\zeta}\:d^3\vec{\xi}\:|\psi^\cZ(\vec{\zeta})|^2 
f(\vec{0},\vec{\xi}) \: |\psi_i^{\cX\cS}(\vec{\xi}+\vec{\zeta})|^2,\hspace*{5mm}i=1,\ldots,M,
\label{Wbiti}
\ee
where $\vec{0}=(0,0,0)$\/. 
Expressions (\ref{Wb0ti}) and (\ref{Wbiti})
are precisely the values obtained in the usual formalism---i..e, disregarding the
translational degrees of freedom $\cZ$\/ of the measuring device---for the probabilities of 
obtaining the given 
outcomes upon measuring a classical statistical mixture (proper mixture---see Sec. 7.2 of Ref. 18) \zzz 
of quantum systems, with the (continuous) elements of  the mixture indexed by $\vec{\zeta}$\/ and having the classical probability density  $p(\vec{\zeta})=|\psi^\cZ(\vec{\zeta})|^2$\/.

To see this, we redo the analysis that led to (\ref{Wb0ti}), (\ref{Wbiti}), leaving out
the observer translational degrees of freedom $\cZ$\/ and taking the initial state to be a proper mixture. The state space is spanned by tensor products of the basis vectors (\ref{basisS}),
(\ref{basisO}), and (\ref{cXivec}) (but not (\ref{cZivec})).  Define the $\cX$\/-space
projection operator
\be
\wh{P}^{\cX}_g=\int\: d^3\vec{\xi} \: g(\vec{\xi})\:\wh{P}^{\cX}_{\vec{\xi}}, \label{PcXg}
\ee
where $g(\vec{\xi})$\/ is a c-number function satisfying
\be
(g(\vec{\xi}))^2=g(\vec{\xi}),\label{gsquared}
\ee
so
\be
g(\vec{\xi})=\mbox{\rm 0 or 1} \hspace*{5mm} \forall \; \vec{\xi}.
\ee
If we choose
\be
g(\vec{\xi})=\theta(a-|\vec{\xi}|), \hspace*{5mm} a>0,\label{g_def}
\ee
i.e.,
\be
g(\vec{\xi})=f(\vec{0},\vec{\xi}).\label{g_eq_f_0}
\ee
then $\wh{P}^{\cX}_g$\/ is  the projector into $\cX$\/-states within
distance $a$\/ of the origin,
\be
\begin{array}{rll}
\wh{P}_g^{\cX}|\cX;\vec{\xi}\ra&=|\cX;\vec{\xi}\ra,
\hspace*{5mm}& |\vec{\xi}| < a,\\
\mbox{}&=0, \hspace*{2.3cm}& |\vec{\xi}| > a.
\end{array} \label{PcXproj}
\ee
To model a finite-range measurement, we take the Hamiltonian\footnote{Subsequent to the completion of this work I became aware of Sec. III of Ref. 23, \zzz 
where essentially the
same  Hamiltonian as in (\ref{HFpr}) appears.}  to be
\be
\wh{H}_{F}\pr=\wh{H}_I \otimes \wh{P}_g^\cX, \label{HFpr}
\ee
\xxx{HFpr}
where $\wh{H}_I$\/ is as given in (\ref{HI}). The time-evolution
operator from the initial time $t_{in}$\/ to time $t$\/,
\be
\wh{U}_{F}\pr=\exp(-i\wh{H}_{F}\pr(t-t_{in})),
\ee
has the form
\be
\wh{U}_{F}\pr=\wh{P}_{\wti{g}}^{\cX}+\sum_{i=1}^M\wh{u}_i^{\cO}\otimes
\wh{P}_i^{\cS}\otimes\wh{P}_g^{\cX}.\label{UFprform}
\ee
Here
\be
\wti{g}(\vec{\xi})=1- g(\vec{\xi}),\label{tilde_g}
\ee
i.e.,
\be
\wti{g}(\vec{\xi})=\wti{f}(\vec{0},\vec{\zeta}), \label{gtilde_eq_ftilde_0}
\ee
so
\be
\wh{P}_{\wti{g}}= \int d^3\vec{\xi}\; \wti{g}(\vec{\xi})\;
	 \wh{P}_{\vec{\xi}}^{\cX}, \label{PftildecX}
\ee
$\wh{P}_{\wti{g}}^{\cX}$\/ is the projector into the subspace of states in $\cX$\/
which are farther from the origin  than $a$\/:
\be
\begin{array}{rll}
\wh{P}_{\wti{g}}^{\cX}|\cX;\vec{\xi}\ra&=|\cX;\vec{\xi}\ra,
\hspace*{5mm}& |\vec{\xi}| > a,\\
\mbox{}&=0, \hspace*{2.3cm}& |\vec{\xi}| < a.
\end{array} \label{PcXtildeproj}
\ee
satisfying, from (\ref{PxicX}), (\ref{cXorthog}), (\ref{PcXg}), (\ref{gsquared}) and 
(\ref{tilde_g})-(\ref{PftildecX}),
\be
\left(\wh{P}_{\wti{g}}^{\cX} \right)^2=\wh{P}_{\wti{g}}^{\cX},
\label{PcXtiledgsquared}
\ee
\be
\wh{P}_{\wti{g}}^{\cX}=1-\wh{P}_g^{\cX},
\ee
\be
\wh{P}_{\wti{g}}^{\cX}\wh{P}_g^{\cX}=\wh{P}_g^{\cX}\wh{P}_{\wti{g}}^{\cX}=0.
\label{PcXtiledgPcXg}
\ee
Using 
(\ref{PcSortho}),   (\ref{bidef}),   (\ref{b_0}),
(\ref{UFprform}), (\ref{PcXtiledgsquared}) and   (\ref{PcXtiledgPcXg}),
the measurement-result operator $\wh{b}$\/ has, at time $t$\/, the value
\be
\wh{b}\pr(t)=\wh{b}_0\otimes\wh{P}^{\cX}_{\wti{g}} 
+ \sum_{i=1}^M   \wh{b}_i  \otimes  \wh{P}_i^\cS   \otimes \wh{P}_g^\cX.\label{b_pr_t}
\ee
($\wh{a}$\/ and $\wh{\vec{x}}$\/ are unchanged by the measurement:
\be
\wh{a}\pr(t)={\wh{U}_{F}}^{\prime\dag}\;\wh{a}\;{\wh{U}}_{F}\pr=\wh{a},
\ee
\be
\wh{\vec{x}}\pr(t)={\wh{U}_{F}}^{\prime\dag}\;\wh{\vec{x}}\;{\wh{U}}_{F}\pr=\wh{\vec{x}}.)
\ee

We take the initial (pre-measurement) state to be a proper mixture  
of 
the vectors 
\be
|{\psi_1}\pr(t_{in}),\vec{\zeta}\ra=|\cO;\bet_0\ra
\sum_{i=1}^M\int d^3\vec{\xi}\: \psi_i^{\cX\cS}(\vec{\xi}+\vec{\zeta})
|\cX;\vec{\xi}\ra |\cS;\al_i\ra, \hspace*{5mm} -\vec{\infty} < \vec{\zeta} <\vec{\infty},
\label{initstatezeta}
\ee
\xxx{initstatezeta}
with classical probability density 
\be
p(\vec{\zeta}) \ge 0,\hspace*{5mm} -\vec{\infty} < \vec{\zeta} <\vec{\infty},
\ee
\be
\int d^3 \vec{\zeta} \: p(\vec{\zeta})=1. \label{probnorm}
\ee

Were the initial state one of the pure states $|{\psi_1}\pr(t_{in}),\vec{\zeta}\ra$\/ we
would conclude from (\ref{b_i_eigen2}), (\ref{b_pr_t}), (\ref{initstatezeta}) and interpretive rule 1
that the degree of freedom $\cO$\/, initially in a state of ignorance at time $t_{in}$\/,
has at the post-measurement time $t$\/ split into $M+1$\/ Everett copies which have respectively obtained measurement results $\bet_i$\/, $i=0,\ldots,M$\/.  One can regard the formalism of a proper mixture as a convenient notation for calculating the results of experiments performed on ordinary classical ensembles of quantum systems each one of which is a pure state.\footnote{This is 
not to say  that any mixture can be regarded in a {\em unique}\/ manner an ensemble of pure states; that  statement would be false. See, e.g., Sec. 2.3 of Ref. 24.\zzz
}
We can then conclude that in each of the quantum systems in the ensemble, the degree of
freedom $\cO$\/ has split into Everett copies as described above.  If, on the other hand, we wish to consider that a single quantum system can be described by  a density operator $\wh{\rho}(t_{in})$\/ which is not a pure state, i.e., one for which
\be
(\wh{\rho}(t_{in}))^2 \neq \wh{\rho}(t_{in}),
\ee
then we must extend in a straightforward manner the initial-state requirements (\ref{initialHPstate}) and 
(\ref{b_i_eigenvalue}) to accommodate
the case of general (not necessarily pure) states.  The initial density operator must be of the 
form
\be
\wh{\rho}(t_{in})=\wh{\rho}_{ obs}(t_{in})\otimes\wh{\rho}_{sys}(t_{in}),\label{initrho}
\ee
\xxx{initrho}
where $\wh{\rho}_{obs}(t_{in})$\/ and $\wh{\rho}_{sys}(t_{in})$\/ act only in the
state spaces of the observer and system, respectively, and where
\be
\wh{b}_i\;\wh{\rho}_{obs}(t_{in})=\bet_i\;\wh{\rho}_{obs}(t_{in}),\hspace*{5mm}i=0,\ldots,M.
\label{beigenrho}
\ee
\xxx{beigenrho}

For the mixture under consideration here,
the density operator 
is
\be
\wh{\rho}\pr(t_{in})=\int d^3\vec{\zeta}\: p(\vec{\zeta})\: 
|\psi\pr(t_{in}),\vec{\zeta}\ra \la\psi\pr(t_{in}),\vec{\zeta}|. \label{density_op_here}
\ee
or, using (\ref{initstatezeta}),
\be
\wh{\rho}\pr(t_{in})=\wh{\rho}\pr_{ obs}(t_{in})\otimes\wh{\rho}\pr_{sys}(t_{in}),
\label{initrhopr}
\ee
\xxx{initrhopr}
where
\be
\wh{\rho}\pr_{ obs}(t_{in})=|\cO;\bet_0\ra\la\cO;\bet_0|, \label{rhoprobs}
\ee
\xxx{rhoprobs}
and
\be
\wh{\rho}\pr_{sys}(t_{in})=\int\; d^3\vec{\zeta}\; p(\vec{\zeta})\;\sum_{i,i=1}^M\;
\int \; d^3\vec{\xi} \; d^3\vec{\xi\pr}\; {\psi_i^{\cX\cS}}^{\ast}(\vec{\xi}+\vec{\zeta})
{\psi_j^{\cX\cS}}(\vec{\xi\pr}+\vec{\zeta})
|\cX;\vec{\xi}\ra|\cS;\al_i\ra\la\cX;\vec{\xi\pr}|\la\cS;\al_j|.
\ee
Using  (\ref{b_i_eigenvalue}) and (\ref{rhoprobs}), 
\be
\wh{b}_i\;\wh{\rho}_{obs}\pr(t_{in})=\bet_i\;\wh{\rho}_{obs}\pr(t_{in}),\hspace*{5mm}i=0,\ldots,M,
\ee

Whichever way we choose to think about proper mixtures, the probabilities for the respective measurement outcomes  are given by
the weights 
\be
W\pr_{b,0}=\int d^3 \vec{\zeta}\: p(\vec{\zeta}) \: W\pr_{b,0,\vec{\zeta}}\;,\label{Wb0zeta}
\ee
for ``no 
measurement made,'' and 
\be
W\pr_{b,i}=\int d^3 \vec{\zeta}\: p(\vec{\zeta}) \: W\pr_{b,i,\vec{\zeta}}\;,
\hspace*{5mm}i=1,\ldots,M,
\label{Wbizeta}
\ee
for ``outcome $\bet_i$\/ measured," $i=1,\ldots,M$\/, where
\be
W\pr_{b,0,\vec{\zeta}}=
\la\psi\pr(t_{in}),\vec{\zeta}|\wh{P}_{\wti{g}}^{\cX} |\psi\pr(t_{in}),\vec{\zeta}\ra,
\ee
\be
W\pr_{b,i,\vec{\zeta}}=
\la\psi\pr(t_{in}),\vec{\zeta}|\wh{P}_i^{\cS} \otimes \wh{P}_{g}^{\cX} |\psi\pr(t_{in}),\vec{\zeta}\ra,\hspace*{5mm} i=1,\ldots,M.
\ee
Using (\ref{Sorthon}), (\ref{Sproj}), (\ref{PcXg}), (\ref{PftildecX}) and (\ref{initstatezeta}), 
we find
\be
W\pr_{b,0}=\sum_{i=1}^M \; \int \; d^3 \vec{\zeta} \; d^3\vec{\xi} \; p(\zeta) \; \wti{g}(\xi)
\; |\psi_i^{\cX\cS}(\vec{\xi}+\vec{\zeta})|^2, \label{Wprb0final}
\ee
\be
W\pr_{b,i}=\int \; d^3 \vec{\zeta} \; d^3 \vec{\xi} \; p(\zeta) \;
g(\vec{\xi}) \; |\psi_i^{\cX\cS}(\vec{\xi}+\vec{\zeta})|, \hspace*{5mm} i=1,\ldots,M.
\ee
Using (\ref{g_eq_f_0}) and (\ref{gtilde_eq_ftilde_0}), we see that these probabilities are identical to the corresponding probabilities (\ref{Wb0ti}),
(\ref{Wbiti}), with the identification 
\be
p(\vec{\zeta})=|\psi^{\cZ}(\vec{\zeta})|^2.
\ee

We conclude that, when a measuring device $\cO$\/ performs a measurement on a system $\cS$\/
by means of the finite-range interaction (\ref{HF}),
no ambiguity is introduced by taking into account the translational degrees of freedom $\cZ$\/
associated with $\cO$\/, either in the possible outcomes or in their respective probabilities.
This is true regardless of the manner in which  the initial-time wave function depends on the $\cZ$\/ degrees of freedom. The presence of this $\cZ$\/-dependence affects the measurement probabilities in the same way that
an initial mixed state for the measured system would in the absence of the $\cZ$\/
degrees of freedom.

\section{Multiple Observers and Spatial Localization}
\label{ConsistencyMultipleSpatial}
\xxx{ConsistencyMultipleSpatial}

We have seen in Sec. \ref{Translation} above that inclusion of spatial degrees of
freedom does not prevent unambiguous determination of outcomes or their associated
probabilities in the  Everett interpretation.  However,
in addition to obtaining definite results for measurements, with corresponding probabilities
consistent with those predicted by quantum mechanics, it is a property
of measuring devices that they can be considered to have well-defined spatial
locations.  In this section we present a scenario involving measurements by multiple observers
 illustrating that 
this property of localization is  present in  the Heisenberg-picture Everett formalism, 
at least when the 
interactions responsible for the measurements are localized  along  the lines of
Sec. \ref{Translation}. 

The measurement scenario is of the following form: Two observers $\cO^{(1)}$\/
and $\cO^{(2)}$\/ each perform localized measurements on a system $\cS$\/.  
By virtue of the fact that these
observers are localized in space, we expect the results of the paired measurements
to be correlated in certain ways.  To see if this is the case, a third observer
$\cG$\/ performs measurements on both $\cO^{(1)}$\/ and $\cO^{(2)}$\/.  (In Everett quantum
mechanics, we must  always consider an explicit
measurement of  the results of measurements made by different observers if we wish to
speak of 
correlations between those results. In the absence of such an additional measurement, we can only
say, e.g., that $\cO^{(1)}$\/ and $\cO^{(2)}$\/ have each split into their respective Everett copies, but cannot talk about correlations between the copies.)

Specifically, the observers $\cO^{(p)}$\/, $p=1,2$\/  wish to examine whether they are in agreement
as to the spatial locations of objects around them. 
They each make use of measuring devices of the type described in Sec. \ref{Translation},
which determine the state of system $\cS$\/ provided $\cS$\/ is located within a distance $a^{(p)}$\/
of the origin of $\cO^{(p)}$\/'s  coordinate system.  The two observers first have
to arrange to use a common coordinate system so that they can meaningfully compare their 
results.  They do so by mounting both measuring
devices on a common platform, so as to maintain a fixed displacement between the devices.
(For simplicity we will assume that the orientation of the platform is fixed.) Let 
the measuring devices be adjusted  so that $\cO^{(p)}$\/ can determine the state of
$\cS$\/ if $\cS$\/ is within distance $a^{(p)}$\/ of the location $\vec{d}^{(p)}$\/, where
$\vec{d}^{(p)}$\/ is measured from the common origin of coordinates fixed in
the platform.  After the $\cO^{(p)}$\/'s have performed their measurements, an observer
$\cG$\/ measures their respective states of awareness. 
We take the interaction by means of which
$\cG$\/ measures the $\cO^{(p)}$\/'s to be such that, at the conclusion of the
measurement process,  $\cG$\/  will with certainty know the state of the $\cO^{(p)}$\/'s.
(E.g., the $\cG$\/ measuring apparatus   
might be mounted on the same platform as $\cO^{(p)}$\/'s in 
such a way that both of the $\cO^{(p)}$\/'s are within range.)

In other words, all three observers are here regarded as a single ``bound structure.''
The spatial location of this structure, in addition to that of the spatial location of the
to-be-measured system $\cS$\/, is a quantum observable and, depending on
the nature of the quantum state vector, is in general  ``smeared out.''
Is this state of affairs consistent with  our usual notions of localization in space?
In the present scenario, these notions lead us to expect the following results to be true:
\begin{description}
\item[Case 1.] If $|\vec{d}^{(1)}-\vec{d}^{(2)}|  
> a^{(1)} + a^{(2)}$\/---i.e., if the regions in which the two measuring devices can interact with 
$\cS$\/ have no overlap---then $\cG$\/ will never find that both $\cO^{(1)}$\/
and $\cO^{(2)}$\/  have determined the state of $\cS$\/.
\item[Case 2.] If, on the other hand,  $\vec{d}^{(1)}=\vec{d}^{(2)}$\/ and  
$ a^{(1)} = a^{(2)}$\/, so the two interaction regions coincide, then $\cG$\/ will
never find that only one of the $\cO^{(p)}$\/'s has determined the state of $\cS$\/.
Rather, $\cG$\/ will always find either that neither has, or that both have, with
the results obtained in the latter case  by $\cO^{(1)}$\/ and $\cO^{(2)}$\/ always being in agreement.
\end{description}

\subsection{State Spaces}

The state space of the observed system is the same  as that employed in Sec. \ref{Translation}, 
its location corresponding to the operator $\wh{\vec{x}}$\/
and its internal state to the operator $\wh{a}$\/.    
The spatial location of the observers is described by a single spatial
coordinate which we will take to be the $\wh{\vec{z}}$\/ of Sec. \ref{Translationmodelstate}.
In other words, the continuous degrees of freedom are the same as in Sec. \ref{Translation}.

The
operators $\wh{b}\pup$\/ 
represent  
the 
states of awareness of the two observers  who measure $\cS$\/:
\be
\wh{b}\pup|\cO\pup;\bet_i\ra=\bet_i|\cO\pup;\bet_i\ra,\hspace*{5mm}i=0,\ldots,M,
\hspace*{5mm}p=1,2,
\ee
with the $\bet_i$\/ nondegenerate (see (\ref{bnondegen})) and the $|\cO\pup;\bet_i\ra$\/
orthonormal,
\be
\la \cO\pup;\bet_i|\cO\pup;\bet_j \ra = \delta_{ij},\hspace*{.5cm} i,j=O,\ldots,M,
\hspace*{.5cm}p=1,2.  \label{Oporthon}
\ee
\xxx{Oporthon}
The state of awareness of $\cG$\/, the observer who measures the respective states of awareness
of $\cO^{(1)}$\/ and $\cO^{(2)}$\/ after they've measured $\cS$\/, is given by
the eigenvalues $\gamma_I$\/ of the operator $\wh{g}$\/, with eigenvectors $|\cG;\gamma_I\ra$\/,
\be
\wh{g}|\cG;\gamma_I\ra=\gamma_I|\cG;\gamma_I\ra,\hspace*{5mm}I=0,1,\ldots,(M+1)^2.
\ee
$I=0$\/ indicates the state of ignorance, while the remaining $(M+1)^2$\/ possible
values of $I$\/ correspond to $\cG$\/ having determined that $\cO^{(1)}$\/ has 
measured $\bet_i$\/ and that $\cO^{(2)}$\/ has measured $\bet_j$\/, $i,j=0,\ldots,M$\/.
That is, there is a mapping $I(i,j)$\/  from the pairs of states $|\cO^{(1)};\bet_i\ra|\cO^{(2)};\bet_j\ra$\/
which $\cG$\/ observes to the $(M+1)^2$\/ non-ignorant states of awareness of
$\cG$\/.

\subsection{Measurement Interactions}  

Define
\bea
f\ppr_{(p)}(\vec{\zeta},\vec{\xi})&=&\theta(a^{(p)}-|\vec{\xi}-(\vec{\zeta}+\vec{d}_{(p)})|),
\hspace*{5mm}a^{(p)}>0,\hspace*{5mm}p=1,2, \label{fpprpdef}\\ 
\wti{f}\ppr_{(p)}(\vec{\zeta},\vec{\xi})&=&1-f\ppr_{(p)}(\vec{\zeta},\vec{\xi}),
\hspace*{5mm}p=1,2,\label{ftildepprpdef}\\
\wh{P}^{\cZ\cX}_{f\ppr(p)}&=&\int  d^3\vec{\zeta}\;d^3\vec{\xi}\;
\wh{P}_{\vec{\zeta}}^{\cZ}\;\wh{P}_{\vec{\xi}}^{\cX}f\ppr_{(p)}(\vec{\zeta},\vec{\xi}),
\hspace*{5mm}p=1,2,\\
\wh{P}^{\cZ\cX}_{\wti{f}\ppr(p)}&=&\int  d^3\vec{\zeta}\;d^3\vec{\xi}\;
\wh{P}_{\vec{\zeta}}^{\cZ}\;\wh{P}_{\vec{\xi}}^{\cX}\wti{f}\ppr_{(p)}(\vec{\zeta},\vec{\xi})
=1-\wh{P}_{f\ppr(p)},\hspace*{5mm}p=1,2.\label{Pfpprp}
\eea
\xxx{1st eq in above array---fpprpdef}
\xxx{2nd eq in above array---ftildepprpdef}
\xxx{4th eq in above array---Pfpprp}
The time evolution operators corresponding to the measurement of $\cS$\/
by the $\cO^{(p)}$\/'s are then
\be
\wh{U}^{(p)\prpr}_F=\wh{P}^{\cZ\cX}_{\wti{f}\ppr(p)}+
\sum_{i=1}^M\wh{u}_i^{\cO(p)}\otimes \wh{P}_i^{\cS} \otimes \wh{P}_{f\ppr(p)}^{\cZ\cX},
\hspace*{5mm}p=1,2.
\ee
obtained from the Hamiltonians 
\be
\wh{H}^{(p)\prpr}_F=\sum_{i=1}^M\wh{h}_i^{\cO(p)}\otimes \wh{P}_i^{\cS} \otimes \wh{P}_{f\ppr(p)}^{\cZ\cX}, \hspace*{5mm}p=1,2,\label{HpFprpr}
\ee
\xxx{HpFprpr}
where
\be
\wh{U}^{(p)\prpr}_F=\exp\left(-i\tau\ppr\wh{H}^{(p)\prpr}_F\right),\hspace*{5mm}p=1,2.
\label{UpFprprHpFprpr}
\ee
\xxx{UpFprprHpFprpr}
$\wh{u}^{\cO(p)}_i$\/ is related to $\wh{h}^{\cO(p)}_i$\/ by 
\be
\wh{u}^{\cO(p)}_i=\exp\left(-i\tau\pup\wh{h}^{\cO(p)}_i\right),
\hspace*{5mm}i=1,\ldots,M,\hspace*{5mm}p=1,2,\label{ucOpi}
\ee
\xxx{ucOpi}
with
\be
\wh{h}^{\cO(p)}_i=i\kappa\pup\left(|\cO\pup;\bet_i\ra\la\cO\pup;\bet_0|-
                              |\cO\pup;\bet_0\ra\la\cO\pup;\bet_i|\right)\hspace*{5mm}p=1,2,
\label{hcOpi}
\ee
\xxx{hcOpi}
and 
\be
\kappa\pup=\frac{\pi}{2\tau\pup},\hspace*{5mm}p=1,2.\label{kappap}
\ee
\xxx{kappap}

The time evolution operator corresponding to the measurement of the $\cO^{(p)}$\/'s
by $\cG$\/ is
\be
\wh{U}^{\cG\prpr}_F=\sum_{i,j=0}^M\wh{u}^{\cG}_{ij}\otimes\wh{P}_i^{\cO(1)}
\otimes\wh{P}_j^{\cO(2)},
\ee
related to the Hamiltonian 
\be
\wh{H}_I^{\cG}=\sum_{i,j=0}^M\wh{h}_{ij}^{\cG}\otimes\wh{P}_i^{\cO(1)}\otimes\wh{P}_j^{\cO(2)}
\label{HcGI}
\ee
\xxx{HcGI}
by
\be
\wh{U}^{\cG\prpr}_F=\exp\left(-i\tau^{\cG}\wh{H}^{\cG}_I\right),\label{UcGprprHcGI}
\ee
\xxx{UcGprprHcGI}
where 
\be
\wh{u}_{ij}^{\cG}=\exp\left(-i\tau^{\cG}\wh{h}_{ij}^{\cG}\right),
\hspace*{5mm}i,j=0,\ldots,M,\label{ucGij}
\ee
\xxx{ucGij}
\be
\wh{h}_{ij}^{\cG}=i \kappa^{\cG}
\left( |\cG; {\gamma}_{I(i,j)}\ra \la \cG; {\gamma}_{0} |-| \cG; {\gamma}_{0} \ra \la \cG; {\gamma}_{I(i,j)} | \right),
\hspace*{5mm}i,j=0,\ldots,M,\label{hcGij}
\ee
\xxx{hcGij}
and
\be
\kappa^{\cG}=\frac{\pi}{2\tau^{\cG}}.\label{kappacG}
\ee
\xxx{kappacG}

\subsection{Time Evolution of Operators}

The operator which generates time evolution from time $t_{in}$\/
to time $t$\/ is 
\be
\wh{U}\ppr_{F2}=\wh{U}^{\cG\prpr}_F\wh{U}^{(2)\prpr}_F\wh{U}^{(1)\prpr}_F,
\label{Utotalprpr}
\ee
\xxx{Utotalprpr}
and the operator corresponding to the state of awareness of $\cG$\/ at time $t$\/ is
\bea
\wh{g}\ppr(t)&=&\wh{U}^{\prpr\dag}_{F2}\;\wh{g}\;\wh{U}\ppr_{F2}\label{gpprt}\\
             &=&\sum_{i,j=0}^M \wh{g}_{ij}\otimes\wh{l}\ppr_{g,ij}\label{gpprtdecomp}
\eea
\xxx{gpprt}
\xxx{gpprtdecomp}
where  
\be
\wh{g}_{ij}=\wh{u}_{ij}^{\cG\dag}\; \wh{g} \; \wh{u}_{ij}^{\cG},\label{gij}
\ee
\xxx{gij}
and
\bea
\wh{l}\ppr_{g,ij}&=&\rule{0cm}{.6cm}
\wh{P}_i^{\cO(1)}\otimes\wh{P}_j^{\cO(2)}\otimes
\wh{P}_{\wti{f}\ppr(1)}^{\cZ\cX}\wh{P}_{\wti{f}\ppr(2)}^{\cZ\cX}\\ \nonumber
&&+\sum_{i\pr=1}^M \left\{\rule{0cm}{.5cm}
\wh{u}_{i\pr}^{\cO(1)\dag}\wh{P}_i^{\cO(1)}\wh{u}_{i\pr}^{\cO(1)}
\otimes\wh{P}_j^{\cO(2)}\otimes\wh{P}_{f\ppr(1)}^{\cZ\cX}\wh{P}_{\wti{f}\ppr(2)}^{\cZ\cX}\right.\\ \nonumber
&&+\wh{P}_i^{\cO(1)}\otimes\wh{u}_{i\pr}^{\cO(2)\dag}\wh{P}_j^{\cO(2)}\wh{u}_{i\pr}^{\cO(2)}
\otimes\wh{P}_{\wti{f}\ppr(1)}^{\cZ\cX}\wh{P}_{f\ppr(2)}^{\cZ\cX}\\ \nonumber
&&\left.+\wh{u}_{i\pr}^{\cO(1)\dag}\wh{P}_i^{\cO(1)}\wh{u}_{i\pr}^{\cO(1)}
\otimes\wh{u}_{i\pr}^{\cO(2)\dag}\wh{P}_j^{\cO(2)}\wh{u}_{i\pr}^{\cO(2)}
\otimes\wh{P}_{f\ppr(1)}^{\cZ\cX}\wh{P}_{f\ppr(2)}^{\cZ\cX}\rule{0cm}{.5cm}\right\}
\otimes\wh{P}_{i\pr}^{\cS}\rule{0cm}{.6cm}, \label{lpprgij}
\eea
\xxx{lpprgij}
satisfying
\be
\wh{l}\ppr_{g,ij}\wh{l}\ppr_{g,i\pr j\pr}=\wh{l}\ppr_{g,ij}\delta_{i i\pr}\delta_{j j\pr}
\label{lgpprijorthon}
\ee
\xxx{lgpprijorthon}
and
\be
\sum_{i,j=0}^M\wh{l}\ppr_{g,ij}=1.
\label{lgpprijcomplete}
\ee
\xxx{lgpprijorthon}

(The operators $\wh{b}\pup$\/ undergo Everett splitting, while $\wh{a}$\/, $\wh{\vec{x}}$\/ and
$\wh{\vec{z}}$\/ are unchanged:
\bea
\wh{b}\pup(t)&=&\wh{U}^{\prpr\dag}_F\;\wh{b}\pup\;\wh{U}^{\prpr}_F
=\wh{b}\pup_0\otimes\wh{P}^{\cZ\cX}_{\wti{f}(p)}\;
+\;\sum_{i=1}^M\;\wh{b}\pup_i\otimes\wh{p}^{\cS}_i\otimes\wh{p}^{\cZ\cX}_{f(p)},\hspace*{5mm}
p=1,2,\\
\wh{a}(t)&=&\wh{U}^{\prpr\dag}_F\;\wh{a}\;\wh{U}^{\prpr}_F=\wh{a},\\
\wh{\vec{x}}(t)&=&\wh{U}^{\prpr\dag}_F\;\wh{\vec{x}}\;\wh{U}^{\prpr}_F=\wh{\vec{x}},\\
\wh{\vec{z}}(t)&=&\wh{U}^{\prpr\dag}_F\;\wh{\vec{z}}\;\wh{U}^{\prpr}_F=\wh{\vec{z}}.)
\eea

\subsection{Initial State}

The observables pertaining to the measured system should be uncorrelated
with those pertaining to the measuring devices. In addition all three
measuring devices should be in their respective states of ignorance. We
therefore take the initial state to be
\be
|\psi\ppr_2(t_{in})\ra=|\cG,\cZ,\cO,t_{in}\ra |\cX,\cS,t_{in}\ra,
\label{psippr2tin}
\ee
\xxx{psippr2tin}
where 
\be
|\cG,\cZ,\cO,t_{in}\ra=\int d^3\vec{\zeta}\;\psi^{\cZ}(\vec{\zeta})|\cZ;\vec{\zeta}\ra|\cG;\gamma_0\ra
|\cO^{(1)};\bet_0\ra|\cO^{(2)};\bet_0\ra
\label{cGcZcOtin}
\ee
\xxx{cGcZcOtin}
and where $|\cX,\cS,t_{in}\ra$\/ is as in eq. (\ref{cXcStin}).
Normalization of $|\psi\ppr_2(t_{in})\ra$\/,
\be\la\psi\ppr_2(t_{in})|\psi\ppr_2(t_{in})\ra=1,
\ee
imposes the constraint
\be
\left( \int \; d^3\vec{\xi}\;\sum_{i=1}^M\;|\psi^{\cX\cS}_i(\vec{\xi})|^2\right)
\;\left(\int \; d^3\vec{\zeta}\;|\psi^{\cZ}(\vec{\zeta})|^2\right)=1.\label{normalization3}
\ee
\xxx{normalization3}

\subsection{Probability}\label{Probability}

The probability that $\cG$\/ determines that the states of awareness
of $\cO^{(1)}$\/ and $\cO^{(2)}$\/ are respectively $\bet_i$\/ and 
$\bet_j$\/ is
\be
W\ppr_{2g,ij}=\la \psi\ppr_2(t_{in})| \wh{l}\ppr_{g,ij}  | \psi\ppr_2(t_{in})\ra,
\label{Wppr2gij}
\ee
\xxx{Wppr2gij}
Using (\ref{cZivec}-\ref{PzetaicZi}), (\ref{cXcStin}), (\ref{fpprpdef}-\ref{Pfpprp}), (\ref{lpprgij}),  (\ref{psippr2tin}) and 
(\ref{cGcZcOtin})
in (\ref{Wppr2gij}),
 we find that
\bea
W\ppr_{2g,ij}&=&\int  \;  \; d^3 \vec{\zeta} \;  d^3\vec{\xi} \;
  |\psi^{\cZ}(\vec{\zeta})|^2 \; 
\nonumber \\
&&\cdot\left[\;\delta_{i0}\;\delta_{j0}\;\wti{f}_{(1)}(\vec{\zeta},\vec{\xi})\;
\wti{f}\ppr_{(2)}(\vec{\zeta},\vec{\xi})\;\sum_{k=1}^M|\psi^{\cX\cS}_k(\vec{\xi})|^2\right. \nonumber\\
&&\;\;\;+(1-\delta_{i0})\;\delta_{j0}\;f\ppr_{(1)}(\vec{\zeta},\vec{\xi})\;
\wti{f}\ppr_{(2)}(\vec{\zeta},\vec{\xi})\;|\psi^{\cX\cS}_i(\vec{\xi})|^2
\nonumber\\
&&\;\;\;+\delta_{i0}\;(1-\delta_{j0})\;\wti{f}\ppr_{(1)}(\vec{\zeta},\vec{\xi})\;
f\ppr_{(2)}(\vec{\zeta}\pr,\vec{\xi})\;|\psi^{\cX\cS}_j(\vec{\xi})|^2
\nonumber\\
&&\;\;\;\left.+\delta_{ij}\;(1-\delta_{i0})\;f\ppr_{(1)}(\vec{\zeta},\vec{\xi})\;
f\ppr_{(2)}(\vec{\zeta},\vec{\xi})\;|\psi^{\cX\cS}_i(\vec{\xi})|^2\rule{0mm}{6mm}\right],
\hspace*{5mm}i,j=0,\ldots,M.
\label{Wppr2g-ij-2}
\eea
\xxx{Wppr2g-ij-2}
Using (\ref{fpprpdef}), (\ref{ftildepprpdef}), (\ref{normalization3}) and  (\ref{Wppr2g-ij-2}) we verify that
\be
\sum_{i,j=0}^M W\ppr_{2g,ij}=1.
\ee

We now consider the two possibilities for the value of the $\vec{d}\pup$\/'s and
the $a^{(p)}$\/'s described above:

\noindent {\bf Case 1.} Suppose that  
\be
|\vec{d}^{(1)}-\vec{d}^{(2)}|  > a^{(1)} + a^{(2)}. \label{bigdsep} 
\ee
\xxx{bigdsep}
Then
\be
f\ppr_{(1)}(\vec{\zeta},\vec{\xi})f\ppr_{(2)}(\vec{\zeta},\vec{\xi})=0 
\hspace{5mm}\forall \; \vec{\zeta},\vec{\xi}.
\label{fprodzero}
\ee
\xxx{fprodzero}

To verify this, assume that, on the contrary,
\be
f\ppr_{(1)}(\vec{\zeta}\pr,\vec{\xi}\pr)f\ppr_{(2)}(\vec{\zeta}\pr,\vec{\xi}\pr)\neq 0 
\nonumber
\ee
for some $\vec{\zeta}\pr$\/, $\vec{\xi}\pr$\/, implying
\bea
f\ppr_{(1)}(\vec{\zeta}\pr,\vec{\xi}\pr) &\neq& 0,\label{fppr1implication1}\\
f\ppr_{(2)}(\vec{\zeta}\pr,\vec{\xi}\pr) &\neq& 0.\label{fppr2implication1}
\eea
\xxx{fppr1implication1}
\xxx{fppr1implication2}
Using the definition (\ref{fpprpdef}), these imply that
\bea
|\vec{\xi}\pr-(\vec{\zeta}\pr+\vec{d}^{(1)})| & < & a^{(1)},\label{fppr1implication2}\\
|\vec{\xi}\pr-(\vec{\zeta}\pr+\vec{d}^{(2)})| & < & a^{(2)},\label{fppr2implication2}
\eea
\xxx{fppr1implication2}
\xxx{fppr2implication2}
so
\be
|\vec{\xi}\pr-(\vec{\zeta}\pr+\vec{d}^{(1)})|+|\vec{\xi}\pr-(\vec{\zeta}\pr+\vec{d}^{(2)})|
< a^{(1)} + a^{(2)}. \label{fpprsumineq}
\ee
\xxx{fpprsumineq}
It follows from the Schwartz inequality that, for any vectors $\vec{A}$\/ and $\vec{B}$\/,
\be
|\vec{A}-\vec{B}| \leq |\vec{A}|+|\vec{B}|, \label{vectorineq}
\ee
\xxx{vectorineq}
(see, e.g., p. 35 of Ref. 25).\zzz  
Therefore (\ref{fpprsumineq}) implies
\be
|\vec{d}^{(1)}-\vec{d}^{(2)}|  \leq  a^{(1)} + a^{(2)},\label{bigdsepcontra}
\ee
\xxx{bigdsepcontra}
contradicting the condition (\ref{bigdsep}).

So, in this case---i.e., ``large'' separation, as given by   
(\ref{bigdsep})---the last term in the square brackets in (\ref{Wppr2g-ij-2}) vanishes,
implying $W\ppr_{2g,ij}=0$\/ unless at least one of the indices $i,j$\/
is zero. That is, there is zero probability that $\cO^{(1)}$\/ and
$\cO^{(2)}$\/ both measure $\cS$\/.

\noindent {\bf Case 2.}   If 
$\vec{d}^{(1)}=\vec{d}^{(2)}$\/ and  
$ a^{(1)} = a^{(2)}$\/ then, from (\ref{fpprpdef}),
\be
f\ppr_{(1)}(\vec{\zeta},\vec{\xi})=f\ppr_{(2)}(\vec{\zeta},\vec{\xi}).\label{fppr1eqfppr2}
\ee
\xxx{fppr1eqfppr2}
From (\ref{fpprpdef}) and (\ref{ftildepprpdef})
\be
f\ppr_{(p)}(\vec{\zeta},\vec{\xi})\wti{f}\ppr_{(p)}(\vec{\zeta},\vec{\xi})=0,
\hspace*{5mm}p=1,2.
\label{fpprftildeppreq0}
\ee
\xxx{fpprftildeppreq0}
Eqs. (\ref{fppr1eqfppr2}) and (\ref{fpprftildeppreq0}) imply that 
the second and third terms in the square brackets in (\ref{Wppr2g-ij-2})
vanish. So, in this case of both $\cO^{(1)}$\/ and $\cO^{(2)}$\/ measuring $\cS$\/
in the same location, $W\ppr_{2g,ij}=0$\/ unless both of the indices $i,j$\/ are
the same.  $\cG$\/ will always observe either that both $\cO^{(1)}$\/ and $\cO^{(2)}$\/
have failed to measure $\cS$\/, or that  both  have measured it and obtained the same value.

\section{Summary and Discussion}\label{Discussion}

When applied to the localized measurement interaction model developed in this paper,
the rules for determining the possible results of measurements and their 
associated probabilities in the Heisenberg-picture Everett interpretation
give unambiguous results  in the presence of continuous translational degrees of freedom
for both measured systems and observers just as they do  in their absence.  
The continuous-spectrum translational degrees of freedom of the measuring
device have the same effect on the measurement results as would a mixed
state for the observed system.

Despite the presence of translational degrees of freedom for 
observers,
the results obtained by multiple observers measuring another
system and subsequently comparing their results will 
accord with what we think of when we say that a macroscopic system like
a measuring device is ``in one place at a time.''  
E.g., if the observation
regions of the observers $\cO^{(1)}$\/
and $\cO^{(1)}$\/ in
Sec. \ref{ConsistencyMultipleSpatial}  are well-separated, they will never both 
measure $\cS$\/; 
if those regions coincide, they will always measure it, or fail to
measure it, in coincidence, and with the same results when they
do measure it.

 To the extent that Stapp's
claim that macroscopic objects, in particular measuring devices,
are ``smeared out over a continuum of locations'' in collapse-free
Everett quantum mechanics is taken to be a statement of underlying
ontology, we have not refuted it here. Rather, we  have argued that such smearing
{\em will not be perceived}\/.  
The examples of Secs. \ref{Translation} and \ref{ConsistencyMultipleSpatial}
indicate that, to the extent that observers operationally determine their
positions in space by measuring the positions  of  objects using
localized measurement interactions and
then comparing their results, each Everett copy of an observer determining
its location in such manner will obtain an unambiguous result, with a precision
dictated by the limitations the apparatus employed.   Thus, 
both internal 
and spatial  properties of macroscopic objects are definite
at what Schlosshauer (Sec. II.B.3 of Ref. 13) \zzz 
terms the ``observational level.''
This is sufficient to put the theory in accord with 
experience; see Sec. II.B.3 of Ref. 13 and Ref. 26.\footnote{Re: probability as an observational-level phenomenon, see Refs. 31-33.\zzz }

The analysis presented in this paper is in the context of a specific finite-range interaction
model.  It is to be expected that the conclusions obtained through this analysis, particularly regarding localization of measuring devices, will carry over to more complicated and realistic measurement models, since these will also be finite-range.  E.g.,  the best models we currently have of
fundamental interactions, quantum field theories,\footnote{The standard model quantum field theory is believed to
correctly describe particles and their interactions down to a length scale of 
$10^{-18}$\/ m.\cite{Gaillardetal99}} involve  basic constituents (operator-valued
fields) which interact only at the same space-time point (see, e.g., Refs. 28 and 29).\zzz

Joos\cite{Joos87} has previously pointed out the importance of variables with discrete spectra for the Everett interpretation, and
Zurek, in a private communication reported by Schlosshauer,  
argues that the core basis problem vanishes at the observational level due to the
approximately-discrete nature of the states of neurons in the human brain: 
``...It is ultimately only in the brain where the perception of denumerability and mutual exclusiveness of events must be accounted for...when  neurons are more appropriately modeled 
as two-state systems, the issue raised by Stapp disappears\ldots'' (Sec. IV.C.1 of Ref. 13).\zzz 

As I have argued in this paper, discrete states for observers
are indeed a necessary ingredient for the resolution of the core basis problem.  
However, 
what is equally important is taking an operational approach to the properties of objects,
including measuring devices.  To obtain information about, say, the location of
a measuring device, one must perform a measurement to determine it, either explictly or
(as in Sec. \ref{ConsistencyMultipleSpatial}) implicitly.  Without such an approach it is 
unclear, e.g.,  how to handle the translational degrees of freedom of the measuring device while it
is being used to measure the property  $\cS$\/ (see Sec. \ref{Translation}). The use of the Heisenberg-picture
Everett formalism makes this operational approach essentially automatic, since the  very definition of when
something has a property (interpretational rule 1) is couched in terms of the time
evolution,  as a result of measurement interaction, of a degree of freedom of a measuring device.

Note that, since we are not employing
the Copenhagen interpretation, there are no classical objects, with c-number locations in space,
present in the formalism with
respect to which to measure the position of quantum objects in any absolute sense.  The only position information which can
be obtained is relative information, as in Sec. \ref{ConsistencyMultipleSpatial}; ``$\cO^{(2)}$\/ is (or is not)
at the same location as $\cO^{(1)}$\/.''

(The
approximately two-valued voltage across a
neuron membrane is certainly sufficient to serve as
an observer-type degree of freedom $\cal{O}$\/, but
it seems unlikely that it is necessary to proceed so
far up the ladder of complexity before encountering a
suitable operator.  Degrees of freedom in inorganic
matter would seem to be adequate, e.g. those related to the excitation
of vibrational  states of a rock upon absorption of a
photon. In the nervous system, the change in the geometry
of a rhodopsin molecule upon absorption of light (see, e.g.,
Ch. 26 of Ref. 34)  
might be another example.\footnote{Rhodopsin, as the site of
interaction between external electromagnetic fields and the
human visual system, has the been the subject of
a variety of proposals, both theoretical and experimental,
regarding quantum measurement.$\mbox{}^{\mbox{\scriptsize(35-39)}}$\/}
Disturbance of the measured system upon measurement,
as when a photon is absorbed, is another, different,
departure from the ideal-measurement model, but a
rather straightforward one
(see, e.g., Sec. 9.2 of Ref. 24) \zzz 
which does not modify
interpretational  issues.)

Several open issues remain.
As noted above (Sec. \ref{ProbabilitySingleObserver}), the derivation of Born-rule 
probability\mbox{}\footnote{Recently van Esch\cite{vanEsch05} 
has  claimed that the Born rule is not derivable from the nonprobabilistic elements of the Everett formalism but 
must rather be an independent
postulate.   He defines, within the usual nonprobabilistic
structure of quantum mechanics with discrete degrees of freedom,  a rule which associates real numbers between
zero and unity with Schr\"{o}dinger-picture Everett branches. He refers to this rule as an ``alternative  probability rule,''  and then argues 
that since this probability rule, which differs from the Born probability rule, can be defined in the context of
the nonprobabilistic elements of Everett quantum theory just as well as the Born rule, the Born rule cannot be derivable
from the nonprobabilistic elements but must instead be an independent postulate.
Van Esch retains in his alternative quantum theory properties equivalent to interpretive rules
1 and 2. 
Therefore, the results of Ref. 16 \zzz  show that, in his alternative quantum theory,
the probability rule would {\em disagree}\/
with the results obtained by a relative-frequency-measuring device in the infinite-ensemble
limit (at least in the physically-interesting case that measuring devices are of finite size
and employ finite amounts of energy in their operation).  It is questionable whether such a rule, {\em bearing no relation to  
long-term relative frequency}\/, should be termed a ``probability'' rule at all.}
from relative frequency in the infinite-ensemble limit given in Ref. 16 \zzz  for systems with
discrete degrees of freedom has not to date been extended to system with continuous
degrees of freedom such as those in the present paper.  
The analysis presented in this paper should also be extended
to include rotational as well as translational degrees of freedom. 
The finite-range interaction Hamiltonian for,
say, a Stern-Gerlach type device measuring the spin of a spin 1/2 particle would differ somewhat from
the form (\ref{HF})---the projection operators (\ref{Sproj}) would be replaced
by projection operators dependent on the orientation of the measuring device,
since  which components of spin are measured depends on this orientation.

The nondegenerate eigenvalues of the $\cO$\/ degree of freedom in any measurement model of the type presented here should correspond to ``facts''---e.g., ``current is flowing in the semiconductor  junction,'' ``light-emitting diode is 'on','' ``the voltage across this  portion of the cell membrane has a certain value''---which are true or false independent of location, orientation or other degrees of freedom of the measuring apparatus.   A genuinely internal isospin-like degree of freedom would of course fit the bill, but more relevant for modeling most measurement-like situations are systems in which the $\cO$\/ degree of freedom is a function of coordinates describing the positions and/or motions of constituent pieces of the measuring apparatus relative to one another.  So, the construction of simple but explicit models in which the $\cO$\/ degree of freedom is constructed in this manner is essential to relate the abstract model presented above to such more realistic measurement devices, and to determine whether these devices too can be described without basis problems in an Everett framework.

Finally, having seen that well-defined isolated measurements can exist in a collapse-free quantum-mechanical model,
we are still left with the question of why  only certain types of
measuring devices, and not others, are essentially always encountered in art and in nature.  
Our instruments and our
brains can detect and distinguish between live cats and dead cats; but
rare indeed are ``Schr\"{o}dinger-cat sensors'' which detect 
superpositions of macroscopically-distinct states of observed systems.\footnote{Rare,
but not nonexistent!\cite{Friedmanetal00}}
The answer to this question is nowadays thought to be decoherence: Due to interaction
with the environment, successive measurements made by such  detectors would  in
all but highly-controlled conditions give {\em uncorrelated}\/ results, and so would have no predictive value. Such detectors
would therefore not be built, nor would observers capable of making such
measurements have evolved (Ref. 5; \zzz see also  Sec. III.E.1 of Ref. 13 \zzz and references therein).
There is no reason to think that 
decoherence cannot continue to play this role\footnote{In addition, decoherence suppresses {\em interference}\/
between possible trajectories of macroscopic objects. Interference is a phenomenon which involves properties of physical systems not just
at individual times but at sequences of times (histories) (see Sec. II of Ref. 42). 
Quantitative analysis of interference suppression  requires  consideration
of the free dynamics of the system between measurement 
interactions  
(Ref. 43; \zzz
 see also Ref. 44 \zzz 
and references therein)
as well as the during-measurement interactions which have been considered  here.
Recently,  suppression of interference through environmental decoherence has been
demonstrated experimentally.$\mbox{}^{\mbox{\scriptsize (45-47)}}$\/ \zzz }
in models of the type presented in this paper, although
of course this issue ought to be examined in detail.
But, at least in the model presented above, decoherence is not required in order for the results of Everett-quantum-mechanical measurements involving measured systems
and measuring devices with discrete and translational degrees of freedom to be unambiguous and compatible with localization of the measuring devices.

\section*{Acknowledgments} 

I would like to thank Jian-Bin Mao, Rainer Plaga, Allen J. Tino and
two anonymous referees for very helpful
discussions and comments.

\section*{Appendix: A Generalized  Operator Expansion Uniqueness Theorem}

In this Appendix we extend the  operator
expansion uniqueness theorem of Sec. 4.2.1 of Ref. 10 \zzz 
to  systems with 
continuous degrees of freedom such as translation, in order to show that the form for $\wh{b}(t)$\/ in 
(\ref{b_t}) is unique.   

\noindent {\bf Operator expansion uniqueness theorem for systems with continuous degrees of freedom:} Let $\wh{b}(t)$\/ be an operator which acts in the product space 
$\cO\otimes\cV$\/ and can be expanded as
\be
\wh{b}(t)=\sum_{i=0}^M \wh{b}\pr_i\otimes \wh{Q}\pr_i. \label{btA}
\ee
\xxx{btA}
Each Hermitian operator $\wh{b}\pr_i$\/, $i=0,\ldots,M$\/ acts nontrivially only in the
$(M+1)$\/-dimensional state space $\cO$\/, and  satisfies 
\be
\wh{b}\pr_i|\cO:0\ra=\bet\pr_i|\cO:0\ra, \hspace*{5mm}i=0,\ldots,M,\label{bpreigenA}
\ee
\xxx{bpreigenA}
\be
\bet\pr_i=\bet\pr_j \Rightarrow i=j,\hspace*{5mm}i,j=0,\ldots,M,\label{bprnondegenA}
\ee
\xxx{bprnondegenA}
for some vector $|\cO:0\ra$\/ in $\cO$\/.
Each Hermitian operator $\wh{Q}\pr_i$\/, $i=0,\ldots,M$\/, acts nontrivially
only in the state space $\cV$\/ which is spanned by basis vectors $|\cV:\vec{l},
\vec{\nu}\ra$\/ carrying
both discrete and continuous indices:
\be
\la\cV:\vec{l},\vec{\nu}|\cV:\vec{l}\pr,\vec{\nu}\pr\ra=\delta_{\vec{l}\;\vec{l}\pr}
\delta(\vec{\nu}-\vec{\nu}\pr), \label{cVorthonormA}
\ee
\xxx{cVorthonormA}
\be
\sum_{\vec{l}}\int d\vec{\nu}|\cV:\vec{l},\vec{\nu}\ra\la\cV:\vec{l},\vec{\nu}|=1.
\label{cVcompleteA}
\ee
\xxx{cVcompleteA}
The $\wh{Q}\pr_i$\/'s form a complete set of nontrivial orthogonal projection operators in $\cV$\/:
\be
\wh{Q}\pr_i\wh{Q}\pr_j=\delta_{ij}\wh{Q}\pr_i,\hspace*{5mm}i=0,\ldots,M,
\label{QprorthoprojA}
\ee
\xxx{QprorthoprojA}
\be
\sum_{i=0}^M\wh{Q}\pr_i=1,\label{QprcompleteprojA}
\ee
\xxx{QprcompleteprojA}
\be
\wh{Q}\pr_i\neq0,\hspace*{5mm}i=0,\ldots,M.\label{QprnonzeroA}
\ee
\xxx{QprnonzeroA}

Suppose that there exists another expansion for $\wh{b}(t)$\/,
\be
\wh{b}(t)=\sum_{i=0}^M \wh{b}\ppr_i\otimes \wh{Q}\ppr_i,
\label{bt2A}
\ee
\xxx{bt2A}
where the $\wh{b}\ppr_i$\/'s are Hermitian operators acting in $\cO$\/ and satisfying
\be
\wh{b}\ppr_i|\cO:0\ra=\bet\ppr_i|\cO:0\ra, \hspace*{5mm}i=0,\ldots,M,\label{bppreigenA}
\ee
\xxx{bppreigenA}
\be
\bet\ppr_i=\bet\ppr_j \Rightarrow i=j,\hspace*{5mm}i,j=0,\ldots,M,\label{bpprnondegenA}
\ee
\xxx{bpprnondegenA}
for the same vector $|\cO:0\ra$\/ in $\cO$\/, and where the $\wh{Q}\ppr_i$\/'s
are Hermitian operators acting in $\cV$\/ satisfying
\be
\wh{Q}\ppr_i\wh{Q}\ppr_j=\delta_{ij}\wh{Q}\ppr_i,\hspace*{5mm}i=0,\ldots,M,
\label{QpprorthoprojA}
\ee
\xxx{QpprorthoprojA}
\be
\sum_{i=0}^M\wh{Q}\ppr_i=1,\label{QpprcompleteprojA}
\ee
\xxx{QpprcompleteprojA}
\be
\wh{Q}\ppr_i\neq0,\hspace*{5mm}i=0,\ldots,M.\label{QpprnonzeroA}
\ee
\xxx{QpprnonzeroA}
Then the $\wh{b}\ppr_i$\/'s and $\wh{Q}\ppr_i$'s are   identical
to the $\wh{b}\pr_i$\/'s and $\wh{Q}\pr_i$'s up to renumbering:
\be
\wh{b}\ppr_i=\wh{b}\pr_{\pi(i)},\hspace*{5mm}i=0,\ldots,M,\label{bpermA}
\ee
\xxx{bpermA}
\be
\wh{Q}\ppr_i=\wh{Q}\pr_{\pi(i)},\hspace*{5mm}i=0,\ldots,M,\label{QpermA}
\ee
\xxx{QpermA}
where $\pi(i)$\/ is some permutation of $i=0,\ldots,M$\/.

\noindent{\bf Proof:} Since the $\wh{Q}\pr_i$\/'s are Hermitian and commute (eq. (\ref{QprorthoprojA})) they have a common set of eigenvectors $|\cV\pr:\vec{l},\vec{\nu}\ra$\/
which span $\cV$\/
and which can be chosen to be orthonormal:
\be
\wh{Q}\pr_i|\cV\pr:\vec{l},\vec{\nu}\ra=
\om\pr_i(\vec{l},\vec{\nu})|\cV\pr:\vec{l},\vec{\nu}\ra,\hspace*{5mm}i=0,\ldots,M,
\label{ompreigenA}
\ee
\xxx{ompreigenA}
\be
\la \cV\pr:\vec{l},\vec{\nu} | \cV\pr:\vec{l}\pr,\vec{\nu}\pr\ra
=\delta_{\vec{l}\;\vec{l}\pr}\delta(\vec{\nu}-\vec{\nu}\pr).\label{cVprorthonormA}
\ee
\xxx{cVprorthonormA}
\be
\sum_{\vec{l}}\int d\vec{\nu}|\cV\pr:\vec{l},\vec{\nu}\ra\la\cV\pr:\vec{l},\vec{\nu}|=1,
\label{cVprcompleteA}
\ee
\xxx{cVprcompleteA}
\be
\wh{Q}\pr_i=\sum_{\vec{l}}\int d\vec{\nu}|\cV\pr:\vec{l},\vec{\nu}\ra\om\pr_i(\vec{l},\vec{\nu})\la\cV\pr:\vec{l},\vec{\nu}|,
\label{QprexpansionA}
\ee
\xxx{QprexpansionA}

Since the $\om\pr_i(\vec{l},\vec{\nu})$\/'s are eigenvalues of projection operators,
\be
\om\pr_i(\vec{l},\vec{\nu})=\mbox{\rm $0$\/ or $1$\/}.
\label{omprvaluesA}
\ee
\xxx{omprvaluesA}

 {\em Lemma 1:}\/ For fixed $\vec{l}$\/, $\vec{\nu}$\/, 
$\om\pr_i(\vec{l},\vec{\nu})=1$\/ for exactly one value of $i$\/;
$\om\pr_i(\vec{l},\vec{\nu})=0$\/ for all other values of $i$\/.

 {\em Proof:}\/ From (\ref{QprcompleteprojA}),
\be
\sum_{i=0}^M\wh{Q}\pr_i|\cV\pr:\vec{l},\vec{\nu}\ra=|\cV\pr:\vec{l},\vec{\nu}\ra,
\label{lemma1-1A}
\ee
\xxx{lemma1-1A}
so, using (\ref{ompreigenA})
\be
\sum_{i=0}^M \om\pr_i(\vec{l},\vec{\nu})|\cV\pr:\vec{l},\vec{\nu}\ra=|\cV\pr:\vec{l},\vec{\nu}\ra.
\label{lemma1-2A}
\ee
\xxx{lemma1-2A}
Using (\ref{cVorthonormA}), 
\be
\sum_{i=0}^M\om\pr_i(\vec{l},\vec{\nu})=1.\label{omprsumto1A}
\ee
\xxx{omprsumto1A}
This can only be consistent with (\ref{omprvaluesA}) if, for any values of
$\vec{l}$\/ and $\vec{\nu}$\/, $\om\pr_i(\vec{l},\vec{\nu})$\/ is equal
to unity for one value of $i$\/ and zero for all the others. 
\rule[-.55mm]{3mm}{3mm}

Similarly, the $\wh{Q}\ppr_i$\/'s have a common complete set of orthonormal eigenvectors:
\be
\wh{Q}\ppr_i|\cV\ppr:\vec{l},\vec{\nu}\ra=
\om\ppr_i(\vec{l},\vec{\nu})|\cV\ppr:\vec{l},\vec{\nu}\ra,\hspace*{5mm}i=0,\ldots,M,
\label{omppreigenA}
\ee
\xxx{omppreigenA}
\be
\la \cV\ppr:\vec{l},\vec{\nu} | \cV\ppr:\vec{l}\pr,\vec{\nu}\pr\ra
=\delta_{\vec{l}\,\vec{l}\pr}\delta(\vec{\nu}-\vec{\nu}\pr),\label{cVpprorthonormA}
\ee
\xxx{cVpprorthonormA}
\be
\sum_{\vec{l}}\int d\vec{\nu}|\cV\ppr:\vec{l},\vec{\nu}\ra\la\cV\ppr:\vec{l},\vec{\nu}|=1,
\label{cVpprcompleteA}
\ee
\xxx{cVpprcompleteA}
\be
\wh{Q}\ppr_i=\sum_{\vec{l}}\int d\vec{\nu}|\cV\ppr:\vec{l},\vec{\nu}\ra\om\ppr_i(\vec{l},\vec{\nu})\la\cV\ppr:\vec{l},\vec{\nu}|,
\label{QpprexpansionA}
\ee
\xxx{QpprexpansionA}
\be
\om\ppr_i(\vec{l},\vec{\nu})=\mbox{\rm $0$\/ or $1$\/},
\label{ompprvaluesA}
\ee
\xxx{ompprvaluesA}
with, for fixed  $\vec{l}$\/ and  $\vec{\nu}$\/, $\om\ppr_i(\vec{l},\vec{\nu})$\/ equal to unity for one value of $i$\/ 
and equal to zero for all the other values of $i$\/.

Define  functions which map from values of $\vec{l}$\/ and $\vec{\nu}$\/
to the subscript indices of the nonzero eigenvalues $\om\pr_i(\vec{l},\vec{\nu})$\/,
$\om\ppr_i(\vec{l},\vec{\nu})$\/:
\be
I\pr(\vec{l},\vec{\nu})=
i\hspace*{5mm}\mbox{\rm s.t.}\hspace*{5mm}\om\pr_i(\vec{l},\vec{\nu})=1,\label{IprdefA}
\ee
\xxx{IprdefA}
\be
I\ppr(\vec{l},\vec{\nu})=
i\hspace*{5mm}\mbox{\rm s.t.}\hspace*{5mm}\om\ppr_i(\vec{l},\vec{\nu})=1.\label{IpprdefA}
\ee
\xxx{IpprdefA}
In other words,
\be
\om\pr_i(\vec{l},\vec{\nu})=\delta_{i,I\pr(\vec{l},\vec{\nu})},\label{omprdelta}
\ee
\xxx{omprdelta}
\be
\om\ppr_i(\vec{l},\vec{\nu})=\delta_{i,I\ppr(\vec{l},\vec{\nu})}.\label{ompprdelta}
\ee
\xxx{ompprdelta}
It will also be useful to define the sets
\be
\cC\pr_i=\{|\cV\pr:\vec{l},\vec{\nu}\ra | I\pr(\vec{l},\vec{\nu})=i\},\hspace*{5mm}i=0,\ldots,M,\label{cCprisetdefA}
\ee
\xxx{cCprisetdefA}
\be
\cC\ppr_i=\{|\cV\ppr:\vec{l},\vec{\nu}\ra | I\ppr(\vec{l},\vec{\nu})=i\},\hspace*{5mm}i=0,\ldots,M.\label{cCpprisetdefA}
\ee
\xxx{cCpprisetdefA}
By Lemma 1,
\be
\{|\cV\pr:\vec{l},\vec{\nu}\ra\}=\bigcup_{i=0}^M\cC\pr_i,\label{cCprunionA}
\ee
\xxx{cCprunionA}
since every ($\vec{l},\vec{\nu}$\/) maps to the unique  $i$\/ for which $\om\pr(\vec{l},\vec{\nu})=1$\/,
\be
\cC\pr_i \bigcap \cC\pr_j=\emptyset,\hspace*{5mm}i \neq j,\label{cCprintersectionA}
\ee
\xxx{cCprintersectionA}
since no  ($\vec{l},\vec{\nu}$\/)  maps to more than one value of $i$\/,
and similarly
\be
\{|\cV\ppr:\vec{l},\vec{\nu}\ra\}=\bigcup_{i=0}^M\cC\ppr_i,\label{cCpprunionA}
\ee
\xxx{cCpprunionA}
\be
\cC\ppr_i \bigcap \cC\ppr_j=\emptyset,\hspace*{5mm}i \neq j,\label{cCpprintersectionA}
\ee
\xxx{cCpprintersectionA}
It is not possible that $\cC\pr_i$\/ is empty for some  value of $i$\/, say $\wti{i}$\/.
That would imply, using (\ref{IprdefA}) and (\ref{cCprisetdefA}), that
there is no $(\vec{l},\vec{\nu})$\/
for which $\om\pr_{\wti{i}}(\vec{l},\vec{\nu})$\/ is nonzero:
$\om\pr_{\wti{i}}(\vec{l},\vec{\nu})=0\hspace*{5mm} \forall \; \vec{l},\vec{\nu}$\/.
With (\ref{QprexpansionA}), this implies that $\wh{Q}\pr_{\wti{i}}=0$\/,
contradicting (\ref{QprnonzeroA}). So,
\be
\cC\pr_i \neq \emptyset,\hspace*{5mm}i=0,\ldots,M,\label{cCprnotempty}
\ee
\xxx{cCprnotempty}
and, similarly,
\be
\cC\ppr_i \neq \emptyset,\hspace*{5mm}i=0,\ldots,M.\label{cCpprnotempty}
\ee
\xxx{cCpprnotempty}

By hypothesis, 
\be
\sum_{i=0}^M \wh{b}\pr_i\otimes \wh{Q}\pr_i=\sum_{i=0}^M \wh{b}\ppr_i\otimes \wh{Q}\ppr_i,
\label{btexpseqA}
\ee
\xxx{btexpseqA}
Taking the matrix elements of both sides of (\ref{btexpseqA}) between
$|\cO:0\ra|\cV\pr:\vec{l},\vec{\nu}\ra$\/ and 
\mbox{$|\cO:0\ra|\cV\pr:\vec{k},\vec{\mu}\ra$\/,} and using (\ref{Oorthon}), (\ref{bpreigenA}),
(\ref{bppreigenA}), (\ref{ompreigenA}), (\ref{cVprorthonormA}), 
(\ref{omppreigenA}), (\ref{cVpprorthonormA}), (\ref{IprdefA}),  (\ref{IpprdefA}),
and Lemma 1,
\be
\left(\bet\pr_{I\pr(\vec{l},\vec{\nu})}-\bet\ppr_{I\ppr(\vec{k},\vec{\mu})}\right)
\la\cV\pr:\vec{l},\vec{\mu}|\cV\ppr:\vec{k},\vec{\mu}\ra=0.\label{betaprminusbetappr1A}
\ee
\xxx{betaprminusbetappr1A}

Fix $(\vec{k},\vec{\mu})=(\vec{k}_1,\vec{\mu}_1)$\/, and define
\be
i\pr_1=I\pr(\vec{k}_1,\vec{\mu}_1).\label{ipr1defA}
\ee
\xxx{ipr1defA}
There must be some $(\vec{l},\vec{\nu})=(\vec{l}_1,\vec{\nu}_1)$\/ such that
\be
\la\cV\pr:\vec{l}_1,\vec{\nu}_1|\cV\ppr:\vec{k}_1,\vec{\mu}_1\ra \neq 0\label{ipneq0A}
\ee
\xxx{ipneq0A}
(If the inner product in (\ref{ipneq0A}) vanished for all $|\cV\pr:\vec{l}_1,\vec{\nu}_1\ra$\/,
that would imply, with (\ref{cVcompleteA}), that $|\cV\ppr:\vec{k}_1,\vec{\mu})\ra=0$,\/
implying $\la\cV\ppr:\vec{k}_1,\vec{\mu}_1)|\cV\ppr:\vec{k}_1,\vec{\mu}_1\ra=0,$\/
 contradicting  (\ref{cVpprorthonormA}).)  
So,
\be
\left(\bet\pr_{i\pr_1}-\bet\ppr_{i\ppr_1}\right)
\la\cV\pr:\vec{l}_1,\vec{\mu}_1|\cV\ppr:\vec{k}_1,\vec{\mu}_1\ra=0,\label{betaprminusbetappr2A}
\ee
\xxx{betaprminusbetappr2A}
where 
\be
{i\ppr_1}=I\ppr(\vec{k},\vec{\mu}),\label{ippr1defA}
\ee
\xxx{ippr1defA}
implies, with (\ref{ipneq0A}), 
\be
\bet\pr_{i\pr_1}=\bet\ppr_{i\ppr_1}.\label{betasequal1A}
\ee
\xxx{betasequal1A}

Consider any other vector $|\cV\pr : \vec{l}_{1b},\vec{\nu}_{1b} \ra$\/  
which has nonzero inner product with \mbox{$|\cV\ppr:\vec{k}_1,\vec{\mu}_1\ra$\/,}
\be
\la \cV\pr : \vec{l}_{1b},\vec{\nu}_{1b}|\cV\ppr:\vec{k}_1,\vec{\mu}_1\ra \neq 0.
\label{ipneq02A}
\ee
\xxx{ipneq02A}
Using (\ref{betaprminusbetappr1A}), (\ref{ipr1defA}) and (\ref{ipneq02A})
\be
\bet\pr_{I\pr(\vec{l}_{1b},\vec{\nu}_{1b})}=\bet\ppr_{i\ppr_1}.\label{betasequal2A}
\ee
\xxx{betasequal2A}
From  (\ref{betasequal1A}) and (\ref{betasequal2A}), 
\be
\bet\pr_{I\pr(\vec{l}_{1b},\vec{\nu}_{1b})}=\bet\pr_{i\pr_1},\label{betasequal3A}
\ee
\xxx{betasequal3A}
which, with the nondegeneracy condition (\ref{bprnondegenA}), implies
\be
I\pr(\vec{l}_{1b},\vec{\nu}_{1b})=i\pr_1.\label{Isequal1A}
\ee
\xxx{Isequal1A}
That is, all vectors $|\cV\pr :\vec{l},\vec{\nu}\ra$\/ which have nonzero inner product
with $|\cV\ppr :\vec{k}_1,\vec{\nu}_1\ra$\/  are in $\cC\pr_{i\pr_1}$\/.
Repeating the same argument with any other vector in $\cC\ppr_{i\ppr_1}$\/ instead
of $|\cV\ppr:\vec{k}_1,\vec{\nu}_1\ra$\/ leads to the
same conclusion. So,
\be
\la\cV\pr : \vec{l},\vec{\nu}|\cV'': \vec{k},\vec{\mu}\ra=
\delta_{i\pr_1,I\pr(\vec{l},\vec{\nu})}
\la\cV\pr : \vec{l},\vec{\nu}|\cV'': \vec{k},\vec{\mu}\ra,\hspace*{5mm}
|\cV'': \vec{k},\vec{\mu}\ra \in \cC\ppr_{i\ppr_1}.\label{ipr1ippr1mappingA}
\ee
\xxx{ipr1ippr1mappingA}

Starting again but this time with a vector in $\cC\ppr_{i\ppr_2}$\/, where
$i\ppr_2 \neq i\ppr_1$\/, we obtain
\be
\la\cV\pr : \vec{l},\vec{\nu}|\cV'': \vec{k},\vec{\mu}\ra=
\delta_{i\pr_2,I\pr(\vec{l},\vec{\nu})}
\la\cV\pr : \vec{l},\vec{\nu}|\cV'': \vec{k},\vec{\mu}\ra,\hspace*{5mm}
|\cV'': \vec{k},\vec{\mu}\ra \in \cC\ppr_{i\ppr_2},\label{ipr2ippr2mappingA}
\ee
\xxx{ipr2ippr2mappingA}
for some $i\pr_2$\/. It's not possible that $i\pr_2=i\pr_1$\/.  
If $|\cV\pr:\vec{l},\vec{\nu}\ra \in \cC\pr_{i\pr_2}$\/,  $|\cV\ppr:\vec{k},\vec{\mu}\ra \in \cC\ppr_{i\ppr_2}$\/,  and 
$\la\cV\pr:\vec{l},\vec{\nu}|\cV\ppr:\vec{k},\vec{\mu}\ra \neq 0$\/, then
we conclude from (\ref{betaprminusbetappr1A})  that
\be
\bet\pr_{i\pr_2}=\bet\ppr_{i\ppr_2}.\label{betasequalprppr2A}
\ee
\xxx{betasequalprppr2A}
If $i\pr_2=i\pr_1$\/ then (\ref{betasequal1A}) and (\ref{betasequalprppr2A}) imply   
\be
\bet\ppr_{i\ppr_1}=\bet\ppr_{i\ppr_2}.
\label{betasequalpr12A}
\ee
\xxx{betasequalpr12A}
But we have chosen $i\ppr_2$\/ to be unequal to  $i\ppr_1$\/
so (\ref{betasequalpr12A}) contradicts the nondegeneracy condition (\ref{bpprnondegenA}).

By virtue of  (\ref{cCprnotempty})
and (\ref{cCpprnotempty}) we can continue in this manner 
to obtain
\bea
\la\cV\pr : \vec{l},\vec{\nu}|\cV'': \vec{k},\vec{\mu}\ra&=&
\delta_{i\pr_1,I\pr(\vec{l},\vec{\nu})}
\la\cV\pr : \vec{l},\vec{\nu}|\cV'': \vec{k},\vec{\mu}\ra,\hspace*{5mm}
|\cV'': \vec{k},\vec{\mu}\ra \in \cC\ppr_{i\ppr_1},\nonumber\\
& \vdots & \nonumber\\
\la\cV\pr : \vec{l},\vec{\nu}|\cV'': \vec{k},\vec{\mu}\ra&=&
\delta_{i\pr_{M+1},I\pr(\vec{l},\vec{\nu})}
\la\cV\pr : \vec{l},\vec{\nu}|\cV'': \vec{k},\vec{\mu}\ra,\hspace*{5mm}
|\cV'': \vec{k},\vec{\mu}\ra \in \cC\ppr_{i\ppr_{M+1}}.\label{ipstructure1A}
\eea
\xxx{ipstructure1A}
where
\be
i\pr_j \neq i\pr_k\hspace*{5mm}\mbox{\rm for}\; \;   j,k=1,\ldots,M+1,\;\;j \neq k\label{ipruniqueA}
\ee
\xxx{ipruniqueA}
\be
i\ppr_j \neq i\ppr_k\hspace*{5mm}\mbox{\rm for} \;\; j,k=1,\ldots,M+1,\;\;j \neq k\label{ippruniqueA}
\ee
\xxx{ippruniqueA}
\be
0 \leq i\pr_j, i\ppr_k, \leq M, \hspace*{5mm}j,k=1,\ldots,M+1.\label{rangeofipripprA}
\ee
\xxx{rangeofipripprA}
Defining the mapping $\pi(i)$\/ by
\be
i\pr_1=\pi(i\ppr_1),i\pr_2=\pi(i\ppr_2),\ldots,i\pr_{M+1}=\pi(i\ppr_{M+1}),\label{mapdefA}
\ee
\xxx{mapdefA}
eqs. (\ref{ipruniqueA})-(\ref{rangeofipripprA}) tell us that $\pi(i)$\/ is a permutation
of $0,\ldots,M$\/. Using (\ref{mapdefA}) we can write (\ref{ipstructure1A}) as
\bea
\la\cV\pr : \vec{l},\vec{\nu}|\cV'': \vec{k},\vec{\mu}\ra&=&
\delta_{\pi(i\ppr_1),I\pr(\vec{l},\vec{\nu})}
\la\cV\pr : \vec{l},\vec{\nu}|\cV'': \vec{k},\vec{\mu}\ra,\hspace*{5mm}
|\cV'': \vec{k},\vec{\mu}\ra \in \cC\ppr_{i\ppr_1},\nonumber\\
& \vdots & \nonumber\\
\la\cV\pr : \vec{l},\vec{\nu}|\cV'': \vec{k},\vec{\mu}\ra&=&
\delta_{\pi(i\ppr_{M+1}),I\pr(\vec{l},\vec{\nu})}
\la\cV\pr : \vec{l},\vec{\nu}|\cV'': \vec{k},\vec{\mu}\ra,\hspace*{5mm}
|\cV'': \vec{k},\vec{\mu}\ra \in \cC\ppr_{i\ppr_{M+1}}.\label{ipstructure2A}
\eea
\xxx{ipstructure2A}
But from (\ref{cCpprisetdefA}),
\be
|\cV'': \vec{k},\vec{\mu}\ra \in \cC\ppr_{i} \Rightarrow i=I\ppr(\vec{k},\vec{\mu}),
\hspace*{5mm} i=0,\ldots,M,
\label{icCirelationA}
\ee
\xxx{icCirelationA}
so, keeping in mind (\ref{cCpprunionA}), the equations (\ref{ipstructure2A}) can be written as
\be
\la\cV\pr : \vec{l},\vec{\nu}|\cV'': \vec{k},\vec{\mu}\ra=
\delta_{ \pi(I\ppr(\vec{k},\vec{\mu})),I\pr(\vec{l},\vec{\nu}) }
\la\cV\pr : \vec{l},\vec{\nu}|\cV'': \vec{k},\vec{\mu}\ra.\label{ipstructure3A}
\ee
\xxx{ipstructure3A}

Using (\ref{cVcompleteA}) and (\ref{QpprexpansionA}), 
\bea
\wh{Q}\ppr_i&=& \sum_{\vec{l}, \vec{l}\pr,\vec{k}}
\int \; d \vec{\nu} \;d \vec{\nu}\pr \;  d \vec{\mu} \; \nonumber\\
&&|\cV\pr:\vec{l},\vec{\nu}\ra\la\cV\pr:\vec{l},\vec{\nu}|\cV\ppr:\vec{k},\vec{\mu}\ra
\om\ppr_i(\vec{k},\vec{\mu})
\la\cV\ppr:\vec{k},\vec{\mu}|\cV\pr:\vec{l}\pr,\vec{\nu}\pr\ra\la\cV\pr:\vec{l}\pr,\vec{\nu}\pr|
\nonumber\\
&=&\sum_{\vec{l}, \vec{l}\pr,\vec{k}}
\int \; d \vec{\nu} \;d \vec{\nu}\pr \;  d \vec{\mu} \;
\delta_{\pi(I\ppr(\vec{k},\vec{\mu})),I\pr(\vec{l},\vec{\nu}))}
\delta_{\pi(I\ppr(\vec{k},\vec{\mu})),I\pr(\vec{l}\pr,\vec{\nu}\pr))} \nonumber \\
&&|\cV\pr:\vec{l},\vec{\nu}\ra\la\cV\pr:\vec{l},\vec{\nu}|\cV\ppr:\vec{k},\vec{\mu}\ra
\om\ppr_i(\vec{k},\vec{\mu})
\la\cV\ppr:\vec{k},\vec{\mu}|\cV\pr:\vec{l}\pr,\vec{\nu}\pr\ra\la\cV\pr:\vec{l}\pr,\vec{\nu}\pr|
\nonumber\\
&=&\sum_{\vec{l}, \vec{l}\pr,\vec{k}}
\int \; d \vec{\nu} \;d \vec{\nu}\pr \;  d \vec{\mu} \;
\delta_{\pi(I\ppr(\vec{k},\vec{\mu})),I\pr(\vec{l},\vec{\nu}))}
\delta_{\pi(I\ppr(\vec{k},\vec{\mu})),I\pr(\vec{l},\pr\vec{\nu}\pr))} 
\delta_{i,I\ppr(\vec{k},\vec{\mu})}
\nonumber \\
&&|\cV\pr:\vec{l},\vec{\nu}\ra\la\cV\pr:\vec{l},\vec{\nu}|\cV\ppr:\vec{k},\vec{\mu}\ra
\la\cV\ppr:\vec{k},\vec{\mu}|\cV\pr:\vec{l}\pr,\vec{\nu}\pr\ra\la\cV\pr:\vec{l}\pr,\vec{\nu}\pr|
\label{Qpprtrans1A}
\eea
\xxx{Qpprtrans1A}
using (\ref{ipstructure3A}) and its complex conjugate, and then
(\ref{ompprdelta}).  So, 
\bea
\wh{Q}\ppr_i&=&\sum_{\vec{l}, \vec{l}\pr,\vec{k}}
\int \; d \vec{\nu} \;d \vec{\nu}\pr \;  d \vec{\mu} \;
\delta_{\pi(i),I\pr(\vec{l}\vec{\nu}))}
\delta_{\pi(i),I\pr(\vec{l}\pr\vec{\nu}\pr))} 
\delta_{i,I\ppr(\vec{k},\vec{\mu})}\nonumber\\
&&|\cV\pr:\vec{l},\vec{\nu}\ra\la\cV\pr:\vec{l},\vec{\nu}|\cV\ppr:\vec{k},\vec{\mu}\ra
\la\cV\ppr:\vec{k},\vec{\mu}|\cV\pr:\vec{l}\pr,\vec{\nu}\pr\ra\la\cV\pr:\vec{l}\pr,\vec{\nu}\pr|.\label{Qppritrans2A}
\eea
\xxx{Qppritrans2A}
Considering the  first  delta function in (\ref{Qppritrans2A}), we see   that the summand/integrand is only
nonzero for values of $(\vec{l},\vec{\nu})$\/ such that 
\be
I\pr(\vec{l},\vec{\nu})=\pi(i).
\label{IprintcondA}
\ee
\xxx{IprintcondA}
However, (\ref{ipstructure3A}) shows that, if (\ref{IprintcondA}) holds,
$\la\cV\pr:\vec{l},\vec{\nu}|\cV\ppr:\vec{k},\vec{\mu}\ra$\/
is only nonzero if 
\be
\pi(I\ppr(\vec{k},\vec{\mu}))=\pi(i),
\ee
or
\be
I\ppr(\vec{k},\vec{\mu})=i \label{IppreqiA}
\ee
\xxx{IppreqiA}
since $\pi(i)$\/ is a permutation.  So the third delta function in
(\ref{Qppritrans2A}) is superfluous:
\bea
\wh{Q}\ppr_i&=&\sum_{\vec{l}, \vec{l}\pr,\vec{k}}
\int \; d \vec{\nu} \;d \vec{\nu}\pr \;  d \vec{\mu} \;
\delta_{\pi(i),I\pr(\vec{l}\vec{\nu}))}
\delta_{\pi(i),I\pr(\vec{l}\pr\vec{\nu}\pr))}\nonumber\\ 
&&|\cV\pr:\vec{l},\vec{\nu}\ra\la\cV\pr:\vec{l},\vec{\nu}|\cV\ppr:\vec{k},\vec{\mu}\ra
\la\cV\ppr:\vec{k},\vec{\mu}|\cV\pr:\vec{l}\pr,\vec{\nu}\pr\ra\la\cV\pr:\vec{l}\pr,\vec{\nu}\pr|
\nonumber\\
&=&\sum_{\vec{l}}\int d\vec{\nu}\;|\cV\pr:\vec{l},\vec{\nu}\ra
\om\pr_{\pi(i)}(\vec{l},\vec{\nu})\la\cV\pr:\vec{l},\vec{\nu}|\label{Qppritrans3A}
\eea
\xxx{Qppritrans3A}
using (\ref{cVprorthonormA}), (\ref{cVpprcompleteA}) and (\ref{ompprdelta}). Comparing the rightmost members
of (\ref{QprexpansionA}) and (\ref{Qppritrans3A}), we conclude
\be
\wh{Q}\ppr_i=\wh{Q}\pr_{\pi(i)},\hspace*{5mm}i=0,\ldots,M ,\label{QED1A}
\ee
\xxx{QED1A}
as claimed.

Using (\ref{QED1A}) in (\ref{btexpseqA})
\be
\sum_{i=1}^M \wh{b}\pr_i\otimes \wh{Q}\pr_i=\sum_{i=1}^M \wh{b}\ppr_i\otimes \wh{Q}\pr_{\pi(i)},
\label{bp2bppr1A}
\ee
\xxx{bp2bppr1A}
Taking the matrix elements of both sides of (\ref{bp2bppr1A}) between arbitrary vectors
$|\cV\pr:\vec{l},\vec{\nu}\ra$\/, $|\cV\pr:\vec{l}\pr,\vec{\nu}\pr\ra$\/, 
using (\ref{cVorthonormA}), (\ref{omppreigenA}) and (\ref{omprdelta}),
summing over $\vec{l}\pr$\/ and integrating over $\vec{\nu}\pr$\/,
\be
\sum_{i=0}^M(\wh{b}\pr_{\pi(i)}-\wh{b}\ppr_i)\delta_{\pi(i),I\pr(\vec{l},\vec{\nu})}
=0.
\label{bp2bppr2A}
\ee
\xxx{bp2bppr2A}
or
\be
\wh{b}\pr_{I\pr(\vec{l},\vec{\nu})}-\wh{b}\ppr_{{\pi}^{-1}(I\pr(\vec{l},\vec{\nu}))}=0.
\label{bp2bppr3A}
\ee
\xxx{bp2bppr3A}
Since $|\cV:\vec{l},\vec{\nu}\ra$\/ is arbitrary and $\cC\pr_i\neq\emptyset$\/
we can conclude that
\be
\wh{b}\pr_i-\wh{b}\ppr_{{\pi}^{-1}(i)}=0,\hspace*{5mm}i=0,\ldots,M,
\label{bp2bppr4A}
\ee
\xxx{bp2bppr4A}
or
\be
\wh{b}\ppr_i=\wh{b}\pr_{\pi(i)},\hspace*{5mm}i=0,\ldots,M,\label{bp2bppr5A}
\ee
\xxx{bp2bppr5A}
as claimed.

\end{document}